\documentclass[aps,prb,reprint,floatfix,groupedaddress,amsmath,amssymb,amsfonts]{revtex4-2}
\usepackage[T1]{fontenc}
\usepackage[varg]{txfonts}
\usepackage[dvipsnames]{xcolor}
\usepackage{physics}
\usepackage{graphicx}
\usepackage{tikz}
\usetikzlibrary{decorations.pathmorphing,decorations.markings}
\usepackage[percent]{overpic}
\usepackage{bm}
\usepackage{bbm}
\usepackage{mathtools}
\usepackage{hyperref}
\usepackage{orcidlink}

\definecolor{cblue}{RGB}{55,126,184}
\hypersetup{colorlinks=true,linkcolor=cblue,citecolor=cblue,urlcolor=cblue} 

\let\geq\geqslant
\let\leq\leqslant

\begin{document}
\title{Magnetic long-range order at finite temperature in two-dimensional hyperbolic lattices}

\author{Alexander Hickey\,\orcidlink{0000-0002-8050-8132}}
\author{Joseph Maciejko\,\orcidlink{0000-0002-6946-1492}}
\affiliation{Department of Physics, University of Alberta, Edmonton, Alberta T6G 2E1, Canada}
\affiliation{Quantum Horizons Alberta \& Theoretical Physics Institute, University of Alberta, Edmonton, Alberta T6G 2E1, Canada}

\begin{abstract}
Infrared singularities of gapless Goldstone modes preclude magnetic long-range order at finite temperature in conventional two-dimensional systems.
By studying the spin-$S$ Heisenberg model on regular tilings of the hyperbolic plane, we show that this obstruction is absent in negatively curved space.
Using spin-wave theory, we find that the zero-energy collective modes required by symmetry carry vanishing local spectral weight and are separated from the thermodynamic bulk magnon continuum by a finite gap in the bulk local spectral density. 
As a result, local transverse correlations remain short ranged, with a finite correlation length, despite the presence of Goldstone modes associated with the broken $\mathrm{SO}(3)$ spin-rotation symmetry. 
Stronger negative curvature is found to suppress quantum fluctuations in bulk thermodynamic quantities, pushing the ordered state toward ``mean-field-like'' behavior.
We further estimate the ordering temperature from the thermal spin-wave correction to the ordered moment.
These results establish hyperbolic geometry as a route to finite-temperature magnetic order that circumvents the Mermin–Wagner obstruction without breaking or modifying the continuous symmetry.
\end{abstract}

\date{\today}
\maketitle
\section{Introduction}\label{sec:introduction}

The stability of magnetic order is controlled not only by symmetry, but also by the geometry that sets the available phase space for fluctuations. 
In the Euclidean Heisenberg model, a state with broken $\mathrm{SO}(3)$ spin-rotation symmetry supports gapless Goldstone modes, and in two spatial dimensions the associated infrared singularity rules out spontaneous magnetic long-range order at finite temperature for short-range interactions~\cite{goldstone1961,goldstone1962,mermin1966,hohenberg1967,coleman1973}. 
This obstruction reflects not only the existence of symmetry-required zero modes, but also the way those modes are embedded in a continuum of long-wavelength excitations. 
Hyperbolic lattices, realized as regular tessellations of the two-dimensional hyperbolic plane, provide a setting in which this connection must be reconsidered.
The question addressed in this work is therefore simple and concrete: In a standard $\mathrm{SO}(3)$-symmetric Heisenberg magnet on a hyperbolic lattice, do the symmetry-required zero modes prevent magnetic order at finite temperature?

This question is part of the broader effort to understand quantum matter in non-Euclidean geometries~\cite{grass2025}, a program that has accelerated with the experimental realization of hyperbolic lattices.
Hyperbolic lattices have emerged as a useful setting for synthetic quantum matter because their connectivity can be engineered independently of an ambient Euclidean crystal~\cite{kollar2019,boettcher2020,bienias2022,zhang2022,lenggenhager2022,chen2023a,huang2024,chen2024a,yuan2025,xu2025}.
At the single-particle level, their spectral and topological properties have been studied extensively using crystallography, hyperbolic band theory, and related real-space approaches~\cite{boettcher2022,maciejko2021,maciejko2022,cheng2022,kienzle2022,attar2022,lenggenhager2023,shankar2024,mosseri2023,lux2023,chen2023,zhang2023a,lux2024}. 
There is also growing interest in hyperbolic matter as platforms for quantum error correction~\cite{breuckmann2016,higgott2024,jahn2021} and holographic simulation~\cite{boyle2020,asaduzzaman2020,basteiro2022,basteiro2023,basteiro2023a,dey2024}. 
Although current experiments have primarily realized single-particle models, proposals based on qubit--photon interactions and programmable atomic or superconducting arrays provide routes toward interacting many-body systems, including spin models~\cite{boettcher2020,bienias2022,ebadi2021,periwal2021,dede2026}.

The many-body physics of hyperbolic lattices is less developed, but several recent results indicate that negative curvature can qualitatively modify familiar condensed-matter phenomena, including Bose--Einstein condensation~\cite{dutkiewicz2026}, charge-density waves~\cite{zhu2021a}, Hubbard antiferromagnetism~\cite{gotz2024}, Dirac mass generation~\cite{roy2024,gluscevich2025}, altermagnetism~\cite{wang2026,petermann2026}, superconductivity~\cite{bashmakov2025,pavliuk2025}, fractional Chern phases~\cite{he2024,he2025,he2025a}, Lieb--Schultz--Mattis constraints~\cite{shankar2026}, and quantum spin liquids~\cite{lenggenhager2025,dusel2025,mosseri2025,vidal2025}.
Despite this progress, the problem of spontaneous symmetry breaking in hyperbolic quantum magnets remains relatively underexplored.

In this work, we consider the spin-$S$ Heisenberg model on regular hyperbolic lattices using spin-wave theory.
While previous work on Heisenberg magnetism in hyperbolic geometries emphasized boundary physics and sublattice-imbalance effects~\cite{changlani2013,gotz2024}, our focus is the thermodynamic bulk response. 
Our approach is to compute the bulk magnon density of states on finite clusters using its continued fraction expansion~\cite{mosseri2023}, thereby suppressing boundary effects.
Within this formulation, we find that the Goldstone modes carry vanishing spectral weight in the thermodynamic limit and are separated from the bulk magnon continuum by a finite spectral gap. 
Consequently, the order parameter remains well defined at finite temperature, circumventing the usual infrared divergence in two-dimensional Euclidean magnets. 
We further find that increasing negative curvature suppresses quantum and thermal spin-wave fluctuations, pushing bulk thermodynamic quantities toward a more mean-field-like regime, while transverse correlations remain short ranged with a finite correlation length.

The remainder of the paper is organized as follows. 
In Sec.~\ref{sec:model_methods} we introduce the nearest-neighbor hyperbolic Heisenberg model and derive the bulk spin-wave spectra for both the ferromagnet and bipartite antiferromagnet from the adjacency spectrum of the underlying graph. 
In Sec.~\ref{sec:results} we present our results and discuss the implications for magnetic ordering at finite temperature and the correlation functions.
We conclude in Sec.~\ref{sec:conclusion} by outlining the main conclusions and discussing possible extensions of this work.

\section{Model \& methods}\label{sec:model_methods}

\subsection{Hyperbolic geometry}\label{subsec:HyperbolicGeometry}

\begin{figure}[t]
\begin{tikzpicture}
  \node[anchor=south west, inner sep=0] (img) at (0,0)
    {\includegraphics[width=\columnwidth]{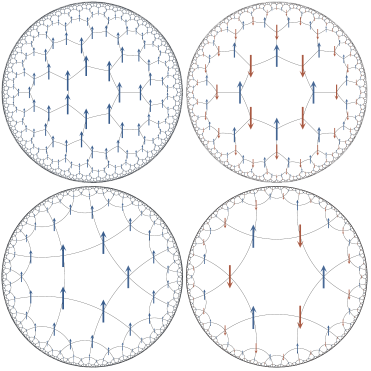}};
  \begin{scope}[x={(img.south east)}, y={(img.north west)}]
    \node[anchor=north west] at (0.000,0.995) {(a)};
    \node[anchor=north west] at (0.490,0.995) {(c)};
    \node[anchor=north west] at (0.000,0.515) {(b)};
    \node[anchor=north west] at (0.490,0.515) {(d)};
    \draw[->, line width=0.45pt] (0.0,0.0) -- (0.1,0.0)
      node[anchor=west, font=\small] {$\vu*{x}$};
    \draw[->, line width=0.45pt] (0.0,0.0) -- (0.0,0.1)
      node[anchor=south, font=\small] {$\vu*{y}$};

  \end{scope}
\end{tikzpicture}
\caption{
Illustration of several hyperbolic $\{p,q\}$ lattices in the Poincaré disk representation of the hyperbolic plane. The (a) $\{7,3\}$ and (b) $\{5,4\}$ lattices are depicted with a fully polarized (ferromagnetic) spin configuration. The (c) $\{8,3\}$ and (d) $\{6,4\}$ lattices are bipartite, and depicted with a N\'eel (antiferromagnetic) spin configuration.
}
\label{fig:hyperbolic-lattices}
\end{figure}
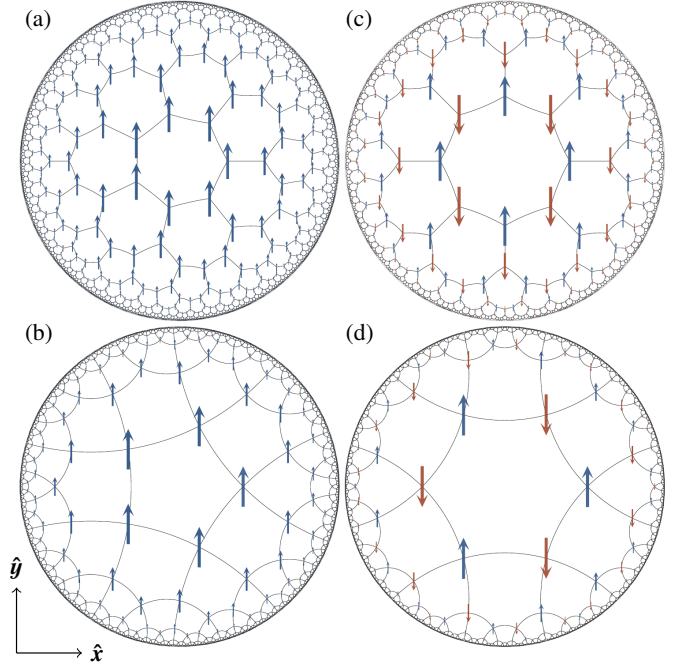

Regular tessellations of a two-dimensional manifold are labeled by the Schl\"afli symbol $\{p,q\}$, where $p$ denotes the number of sides of each polygonal face and $q$ the coordination number, i.e. the number of polygons meeting at each vertex~\cite{coxeter1973,magnus1974}. 
The sign of $(p-2)(q-2)-4$ distinguishes the three constant-curvature geometries: $(p-2)(q-2)<4$ corresponds to spherical tessellations, $(p-2)(q-2)=4$ to the three regular Euclidean tilings~\cite{magnus1974}, and $(p-2)(q-2)>4$ to regular tilings of the hyperbolic plane~\cite{coxeter1998}.
In the following we restrict attention to Euclidean and hyperbolic lattices, while positively curved spherical tessellations lie outside the scope of the present work.
Examples of hyperbolic $\{p,q\}$ lattices considered in this work are shown in Fig.~\ref{fig:hyperbolic-lattices}, where we also indicate the ferromagnetic and bipartite antiferromagnetic classical spin configurations discussed in Sec.~\ref{subsec:HeisenbergModel}.

A convenient conformal representation of the hyperbolic plane is the Poincar\'e disk~\cite{levy1997},
\begin{equation}
    \mathbb{D}=\{z\in\mathbb{C}:|z|<1\},
\end{equation}
equipped with the metric
\begin{equation}
    \dd s^2 = \frac{4}{|\kappa|}\frac{|\dd z|^2}{(1-|z|^2)^2},
    \label{eq:poincare_metric}
\end{equation}
where $\kappa<0$ is the Gaussian curvature.
In this representation, geodesics are either diameters of the disk or circular arcs that intersect the boundary at a right angle~\cite{coxeter1998,magnus1974}. 
The geodesic distance between any two points $z,z'\in\mathbb{D}$ is
\begin{equation}
    r(z,z')
    =
    \frac{1}{\sqrt{|\kappa|}}
    \mathrm{arcosh}\!\left[
        1+\frac{2|z-z'|^2}{(1-|z|^2)(1-|z'|^2)}
    \right].
    \label{eq:geodesic_distance}
\end{equation}
A regular hyperbolic lattice is obtained by placing vertices in $\mathbb{D}$ and connecting nearest neighbors by geodesic segments of equal length, as illustrated in Fig.~\ref{fig:hyperbolic-lattices}.

The local geometry of a regular $\{p,q\}$ tessellation is fixed by the right triangle formed by the center of a $p$-gon, one of its vertices, and the midpoint of one of its edges. 
Its three internal angles are $\pi/p$, $\pi/q$, and $\pi/2$, so hyperbolic trigonometry gives the nearest-neighbor geodesic edge length $a$ as~\cite{mosseri1982}
\begin{equation}
    \cosh \left(\frac{1}{2} \sqrt{|\kappa|a^2}\right)
    =
    \frac{\cos(\pi/p)}{\sin(\pi/q)}.
    \label{eq:pq_curvature_relation}
\end{equation}
Equation~\eqref{eq:pq_curvature_relation} shows that the combination $|\kappa|a^2$ is fixed entirely by the Schl\"afli symbol $\{ p,q \}$. 
Thus, once the microscopic edge length $a$ is specified, the Gaussian curvature of the supporting manifold is no longer an independent parameter: each regular $\{p,q\}$ tiling carries an intrinsic curvature set by its local geometry.

The limit $p\to\infty$ at fixed coordination $q$ corresponds to the $q$-regular Bethe lattice. 
Geometrically, this limit may be viewed as the $\{\infty,q\}$ regular tiling of the hyperbolic plane: the elementary faces become infinite-sided polygons, and the graph contains no closed cycles of finite length~\cite{mosseri1982,soderberg1993}. 
This unifies several Euclidean lattices, finite-$p$ hyperbolic lattices, and the $q$-regular Bethe lattices within a single $\{p,q\}$ family. 
For fixed $q$, increasing $p$ interpolates between the Euclidean lattice (when one exists) and the Bethe lattice limit.

Negative curvature also implies that finite open patches of hyperbolic lattices possess an extensive boundary, in contrast to Euclidean lattices where the fraction of sites on the boundary vanishes in the thermodynamic limit~\cite{sakaniwa2009,baek2009,mosseri2023}.
Closely related finite-size effects, including persistent boundary degree deficits, have also been analyzed recently in hyperbolic circuit realizations~\cite{saa2026}.
This complicates the extraction of bulk observables from finite clusters and must be accounted for when interpreting results from numerical calculations on finite flakes.
Finally, when $p$ is even the lattice is bipartite~\cite{diestel2025}, since the even-sided polygonal faces admit a proper two-coloring of their vertices, permitting a two-sublattice ordered state. 
Odd-$p$ lattices are non-bipartite and frustrate such ordering. 
This distinction will be central in Sec.~\ref{subsec:HeisenbergModel} when discussing the classical ground states of the Heisenberg model.

\subsection{Heisenberg model}\label{subsec:HeisenbergModel}

In the following, we consider the spin-$S$ nearest-neighbor Heisenberg model~\cite{heisenberg1928}
\begin{equation}
    H = J\sum_{\langle i,j\rangle} \vb*{S}_i \cdot \vb*{S}_j,
    \label{eq:HeisenbergModel}
\end{equation}
defined on the regular $\{p,q\}$ lattices introduced in Sec.~\ref{subsec:HyperbolicGeometry}. 
Here $\vb*{S}_i$ is a spin-$S$ operator at site $i$, and the sum runs over nearest-neighbor bonds $\langle i,j\rangle$ of the lattice. 
The Hamiltonian in Eq.~\eqref{eq:HeisenbergModel} is invariant under global $\mathrm{SO}(3)$ spin rotations, so any ordered phase ($\expval{\vb*{S}_i} \neq 0$) is associated with the spontaneous breaking of a continuous symmetry~\cite{anderson1952,goldstone1961,goldstone1962,auerbach1994}.

It is useful to note that the spin directions in Eq.~\eqref{eq:HeisenbergModel} are internal $\mathrm{SO}(3)$ variables, not tangent vectors on the hyperbolic plane.
The scalar product $\vb*{S}_i\cdot\vb*{S}_j$ is therefore taken in a common spin space and does not include a spin connection or any rotation associated with parallel transport on the underlying curved manifold.
Consequently, for the isotropic Heisenberg model considered here, negative curvature affects the spin waves through the graph geometry and spectrum, but does not directly rotate the spin quantization axes from site to site.
This should be contrasted with tangent-vector excitations, such as phonons, whose components live in the tangent bundle and directly sense curvature through holonomy~\cite{sela2025}.
If additional microscopic couplings tie the spin direction to a local lattice or orbital frame, the effective spin Hamiltonian may acquire bond-dependent spin rotations, anisotropies, or Dzyaloshinskii--Moriya-like interactions~\cite{streubel2016}.

The sign of $J$ in Eq.~\eqref{eq:HeisenbergModel} distinguishes the ferromagnetic and antiferromagnetic cases. 
For $J<0$, the exact ground state is the fully polarized ferromagnet, shown schematically in Figs.~\ref{fig:hyperbolic-lattices}(a) and~\ref{fig:hyperbolic-lattices}(b). 
For $J>0$, the classical energy is minimized by antiparallel alignment of neighboring spins. 
On a bipartite lattice, which for regular $\{p,q\}$ tilings requires even $p$, the classical ground state is the two-sublattice N\'eel state also illustrated in Figs.~\ref{fig:hyperbolic-lattices}(c) and~\ref{fig:hyperbolic-lattices}(d). 
In the following, whenever we discuss the antiferromagnet we restrict attention to this bipartite (even $p$) case.
Although it is not itself an exact eigenstate of Eq.~\eqref{eq:HeisenbergModel}, the N\'eel state captures the symmetry-broken order that is known to persist in the thermodynamic limit at $T=0$~\cite{kubo1952,anderson1952,manousakis1991}. 
In Euclidean space, however, the existence of such order at finite temperature is highly constrained. 
In particular, the Mermin--Wagner--Hohenberg theorem~\cite{mermin1966,hohenberg1967} forbids the spontaneous breaking of a continuous symmetry at any finite temperature in spin models with short-range interactions on Euclidean lattices of spatial dimension $d \leq 2$. 
As a consequence, finite-temperature magnetic long-range order is absent in the conventional two-dimensional Euclidean Heisenberg model. 
Ordered ground states and their associated spin-wave excitations nevertheless remain well defined at $T=0$~\cite{kubo1952,anderson1952,manousakis1991,castilla1991,igarashi1992,hamer1992,gochev1993,kubo1988,igarashi2005}.
The Mermin--Wagner--Hohenberg theorem does not carry over to regular hyperbolic lattices.
As we show in Sec.~\ref{sec:results}, the zero-energy modes are separated from the thermodynamic bulk magnon continuum by a finite spectral gap.
This bulk spectral gap removes the finite-temperature spin-wave infrared divergence found on two-dimensional Euclidean lattices.

A further distinction from Euclidean lattices is that finite open hyperbolic clusters retain an extensive boundary, so boundary physics can remain significant even for very large systems~\cite{boettcher2020,mosseri2023}. 
In the present work, however, we are primarily interested in bulk magnetic properties. 
The analysis that follows employs spin-wave theory to characterize the interior, thermodynamic-limit behavior of Eq.~\eqref{eq:HeisenbergModel}, explicitly excluding the boundary-localized features of finite clusters.

\subsection{Spin-wave theory}

To characterize the low-energy subspace of Eq.~\eqref{eq:HeisenbergModel}, we represent the spin-$S$ operators in terms of Holstein--Primakoff bosons~\cite{holstein1940},
\begin{align}
    \vb*{S}_i
    &=
    \left(S-a_i^\dagger a_i^{}\right)\vu*{e}_{i,0}
    + \sqrt{S}\left[
    \left(1-\frac{a_i^\dagger a_i^{}}{2S}\right)^{1/2}
    a_i^{}\,\vu*{e}_{i,-}
    + \mathrm{H.c.}
    \right],
    \label{eq:HP_representation}
\end{align}
where $\vu*{e}_{i,\pm} \equiv ( \vu*{e}_{i,x}\pm i\vu*{e}_{i,y} ) / \sqrt{2}$, and $\{\vu*{e}_{i,x},\vu*{e}_{i,y},\vu*{e}_{i,0}\}$ is a local orthonormal frame satisfying $\vu*{e}_{i,x}\times\vu*{e}_{i,y}=\vu*{e}_{i,0}$. 
The vectors $\vu*{e}_{i,0}$ are chosen along the local classical ordering directions~\cite{kittel1991}.

Due to the $\mathrm{SO}(3)$ symmetry of Eq.~\eqref{eq:HeisenbergModel}, the global orientation of the
ordered state is spontaneous.
Without loss of generality, we choose the ordered moments to lie
along $\vu*{y}$ (see Fig.~\ref{fig:hyperbolic-lattices}), fixing the direction of spontaneous symmetry breaking.
Thus, for the ferromagnet one has $\vu*{e}_{i,0}=\vu*{y}$ on every site, whereas for the bipartite antiferromagnet one may take $\vu*{e}_{i,0}=\pm \vu*{y}$ on the two sublattices.
Equivalently, after a $\pi$ rotation of one sublattice about the global $\vu*{x}$ direction, all spins are aligned with the same local quantization axis, such that a single boson species $a_i$ suffices on every site~\cite{auerbach1994}.
This rotation produces the anomalous terms in Eq.~\eqref{eq:hDelta_AFM}, which are diagonalized by the Bogoliubov transformation in Eq.~\eqref{eq:boson_bogoliubov}.

Expanding Eq.~\eqref{eq:HP_representation} in powers of $1/S$ leads to the series representation of Eq.~\eqref{eq:HeisenbergModel} $H = S^2 \sum_{m=0}^{\infty} S^{-m/2} H_m$.
When the expansion is performed about a classical ground state, the linear term vanishes, $H_1=0$~\cite{kittel1991}, and linear spin-wave theory is obtained by retaining the quadratic term $H_2$, which is of order~$S$.

For a regular $\{p,q\}$ lattice, the linear spin-wave Hamiltonian is entirely determined by the adjacency matrix $\vb*{A}$ of the underlying $q$-regular graph,
defined by $A_{ij}=1$ for nearest-neighbor sites and $A_{ij}=0$ otherwise.
In the site basis it can be written in the generic bosonic Bogoliubov form~\cite{colpa1978,kondo2020}
\begin{equation}
    H
    =
    \mathcal{E}_{\rm cl}
    + \sum_{ij} \left[ \Omega_{ij} a_i^\dagger a_j
    + \frac{1}{2}
    \left(
        \Lambda_{ij} a_i^\dagger a_j^\dagger + \mathrm{H.c.}
    \right)
    \right]
    + O(S^0),
    \label{eq:H2_site_basis}
\end{equation}
where $\mathcal{E}_{\rm cl} = -S\;\text{Tr}[\vb*{\Omega}]/2$ is the classical ground-state energy. The matrices $\vb*{\Omega}$ and
$\vb*{\Lambda}$ take the form
\begin{equation}
    \vb*{\Omega} = |J|S\left(q\mathbbm{1}-\vb*{A}\right),
    \qquad
    \vb*{\Lambda}=0,
    \label{eq:hDelta_FM}
\end{equation}
for the Heisenberg ferromagnet ($J<0$), and
\begin{equation}
    \vb*{\Omega} = JSq\,\mathbbm{1},
    \qquad
    \vb*{\Lambda} = JS\,\vb*{A},
    \label{eq:hDelta_AFM}
\end{equation}
for the bipartite Heisenberg antiferromagnet ($J>0$).

Next, since $\vb*{A}$ is real and symmetric, it admits an orthonormal eigenbasis $\{ \vb*{\phi}_n \}$ with real eigenvalues ${ \lambda_n }$.
For a $q$-regular graph, the uniform Perron--Frobenius vector
\begin{equation}
\vb*{\phi}_0
=
\frac{1}{\sqrt{N}}
\begin{pmatrix}
1 & 1 & \cdots & 1
\end{pmatrix}^{\intercal}
\label{eq:uniform_eigenvector}
\end{equation}
has eigenvalue $+q$.
If the graph is bipartite, the staggered vector with opposite signs on the two sublattices has eigenvalue $-q$.

The corresponding bosonic normal modes are defined as
\begin{equation}
    \alpha_n \equiv \sum_i [\vb*{\phi}_n]_i \,a_i,
\end{equation}
such that Eq.~\eqref{eq:H2_site_basis} factorizes into independent $2\times 2$ Nambu blocks,
\begin{equation}
    H
    =
    \left(1+\frac{1}{S}\right) \mathcal{E}_{\rm cl}
    + \frac{1}{2}\sum_n
    \begin{pmatrix}
        \alpha_n^\dagger & \alpha_n
    \end{pmatrix}
    \vb*{M}_n
    \begin{pmatrix}
        \alpha_n \\
        \alpha_n^\dagger
    \end{pmatrix}
    .
    \label{eq:HP_BdG_general}
\end{equation}
Here, $\vb*{M}_n$ is the $2\times2$ Nambu matrix associated with the adjacency matrix eigenvector $\vb*{\phi}_n$:
\begin{equation}
    \vb*{M}_n =
    |J|S
    \begin{pmatrix}
        q-\lambda_n & 0 \\
        0 & q-\lambda_n
    \end{pmatrix}
\end{equation}
for the ferromagnet, while
\begin{equation}
    \vb*{M}_n =
    JS
    \begin{pmatrix}
        q & \lambda_n \\
        \lambda_n & q
    \end{pmatrix}
\end{equation}
for the bipartite antiferromagnet.
Each block is diagonalized by a bosonic Bogoliubov transformation~\cite{colpa1978,kittel1991}
\begin{equation}
    \gamma_n \equiv u_n \alpha_n + v_n^{} \alpha_n^\dagger,
    \label{eq:boson_bogoliubov}
\end{equation}
with $u_n^2-v_n^2=1$, yielding
\begin{equation}
    H
    =
    \left(1+\frac{1}{S}\right) \mathcal{E}_{\rm cl}
    + \sum_n
    \omega_n \left( \gamma_n^\dagger \gamma_n^{} + \frac{1}{2} \right)
    + O(S^0)
    .
    \label{eq:diagonalized_H}
\end{equation}
For the ferromagnet,
$\vb*{\Lambda}=0$, so the quadratic Hamiltonian is number conserving and $\vb*{M}_n$ is diagonal.
The Bogoliubov transformation in Eq.~\eqref{eq:boson_bogoliubov} therefore reduces to an ordinary unitary transformation, with
\begin{equation}
    \omega_n = |J|S(q-\lambda_n), \quad
    u_n = 1, \quad
    v_n = 0.
    \label{eq:FM_solution}
\end{equation}
For the bipartite antiferromagnet one obtains
\begin{align}
    \omega_n &= JS\sqrt{q^2-\lambda_n^2}, \nonumber \\     
    u_n = \sqrt{\frac{JSq+\omega_n}{2\omega_n}}, &\quad
    v_n = \operatorname{sgn}(\lambda_n)\sqrt{\frac{JSq-\omega_n}{2\omega_n}}.
    \qquad
    \label{eq:AFM_solution}
\end{align}
The spin-wave frequencies $\omega_n$ are real and non-negative for all modes provided $|\lambda_n| \leq q$, which is guaranteed since the spectral radius of the adjacency matrix of a connected $q$-regular graph equals $q$~\cite{brouwer2012}. 
This confirms the stability of the classical ferromagnetic and N\'eel states depicted in Fig.~\ref{fig:hyperbolic-lattices} as a valid expansion point for linear spin-wave theory.
The zero modes ($\omega_n = 0$), which occur at $\lambda_n = q$ for the ferromagnet and $\lambda_n = \pm q$ for the antiferromagnet, correspond to zero-energy Goldstone excitations associated with the spontaneous breaking of $\mathrm{SO}(3)$ spin-rotation symmetry.
Note that throughout this work, we use the term \emph{Goldstone mode} to refer to the zero-energy collective modes generated by the broken global spin-rotation symmetry. 

Equations~\eqref{eq:hDelta_FM}--\eqref{eq:hDelta_AFM} and \eqref{eq:FM_solution}--\eqref{eq:AFM_solution} establish the central result of this section: for both the ferromagnet and the bipartite antiferromagnet, the entire linear spin-wave spectrum is determined by the eigenvalues of the adjacency matrix $\vb*{A}$ of the underlying $q$-regular graph. 
The magnetic problem on a hyperbolic lattice thus reduces to a spectral problem on that lattice's graph.

In practice, although our interest lies in the bulk properties of Eq.~\eqref{eq:HeisenbergModel} in the thermodynamic limit, the adjacency matrices are generated numerically on large finite clusters.
Unlike Euclidean lattices, hyperbolic lattices do not possess a conventional translation symmetry, but instead admit a generalized non-Abelian translation group~\cite{maciejko2021,maciejko2022,lenggenhager2023}. 
In this case, Bloch's theorem does not provide a conventional crystal momentum to label the eigenstates.
This complicates the use of Bloch's theorem to obtain a simple momentum-space description, so we work directly in real space.
To access the corresponding bulk density of states while suppressing spurious boundary contributions, we employ the continued-fraction method~\cite{haydock1972,gaspard1973,haydock1975,viswanath1994}.
In this approach, the local Green's function associated with a site deep in the interior of the cluster is represented as an infinite continued fraction whose recursion coefficients are generated iteratively from moments of the adjacency matrix.
For hyperbolic $\{p,q\}$ lattices, where the coefficients rapidly converge~\cite{mosseri2023}, the resulting continued fraction provides an accurate reconstruction of the thermodynamic bulk local density of states, and hence of the bulk spectral properties of $\vb*{A}$.
More details of this construction are provided in App.~\ref{appendix:LDOS}.

\begin{figure*}[ht!]
    \centering
    \begin{overpic}[width=0.98\textwidth]{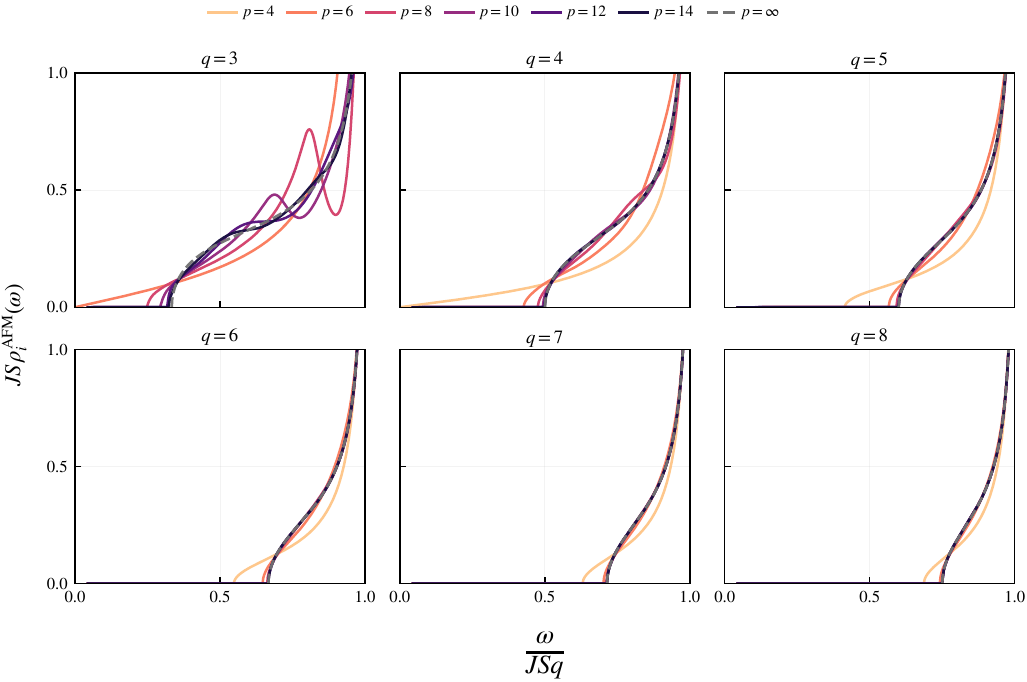}
    \end{overpic}
    \caption{
    Bulk local magnon density of states at $O(S)$ for the bipartite Heisenberg antiferromagnet on Euclidean and hyperbolic $\{p,q\}$ lattices. 
    The bipartite Euclidean lattices, the $\{6,3\}$ honeycomb lattice and the $\{4,4\}$ square lattice, have gapless bulk spectra, whereas the hyperbolic lattices have a finite lower band edge in the thermodynamic bulk local density of states. 
    The grey dashed curves denote the exact solution for the Bethe lattice limit ($p=\infty$).
    }
    \label{fig:AFM_LDOS}
\end{figure*}

The remainder of this paper focuses on the bulk thermodynamic properties of the hyperbolic Heisenberg model. 
We are particularly interested in how the spin-wave spectrum, local response functions, quantum corrections, and finite-temperature ordering scale with the intrinsic Gaussian curvature $|\kappa|a^2$, fixed by Eq.~\eqref{eq:pq_curvature_relation}, of a hyperbolic lattice. 
For a general $\{p,q\}$ lattice, the bulk local density of states must be reconstructed numerically using the continued-fraction method described in App.~\ref{appendix:LDOS}. 
The Bethe lattice, corresponding to the limit $p\to\infty$ at fixed coordination number $q$, admits exact closed-form expressions and is discussed separately in App.~\ref{app:BetheLattice}. 
Together, these results allow us to compare finite-$p$ hyperbolic lattices with an analytically controlled reference point.

\section{Results}\label{sec:results}

\subsection{Bulk density of states and transverse susceptibility}
\label{subsec:local_transverse_susceptibility}

The central numerical object in the following analysis is the bulk local density of states, which serves as the common input for the local response functions and bulk thermodynamic quantities computed throughout this section.
As discussed in App.~\ref{appendix:LDOS}, this quantity is obtained from large finite clusters by applying the continued-fraction method~\cite{haydock1972,haydock1975} to a site chosen far from the boundary of the lattice.  
This yields the thermodynamic bulk local density of adjacency eigenvalues with boundary contributions suppressed.
The spin-wave relations in Eqs.~\eqref{eq:FM_solution} and \eqref{eq:AFM_solution} then map this adjacency spectrum to the corresponding magnon density of states. 
In this way, the density of states is computed first and subsequently used as the input for \emph{local} response functions and thermodynamic quantities.

Let $\{\vb*{\phi}_n\}$ denote the orthonormal eigenvectors of the adjacency matrix $\vb*{A}$, with
$\vb*{A}\vb*{\phi}_n=\lambda_n\vb*{\phi}_n$. 
The local density of adjacency eigenvalues at a bulk site $i$ is
\begin{equation}
    \nu_i(\lambda)
    =
    \sum_n [\vb*{\phi}_n]_i^2 \delta(\lambda-\lambda_n).
    \label{eq:adjacency_ldos}
\end{equation}
The isolated extremal eigenvectors discussed in Sec.~\ref{sec:model_methods} must be distinguished from the thermodynamic bulk local spectral measure.
The uniform Perron--Frobenius vector at $\lambda=q$ has local weight $[\vb*{\phi}_0]_i^2=1/N$ at any fixed site, and the corresponding staggered vector in the bipartite case has the same scaling.
Thus, although these modes give the zero-frequency Goldstone modes of the spin-wave Hamiltonian, their contribution to $\nu_i(\lambda)$ vanishes in the bulk thermodynamic limit $N\rightarrow\infty$.

For the bipartite antiferromagnet ($J>0$), the magnon energy is $ \omega=JS\sqrt{q^2-\lambda^2}$, giving
\begin{equation}
    \rho_i^{\rm AFM}(\omega)
    =
    \frac{2\omega}{J^2 S^2 \sqrt{q^2-\left(\frac{\omega}{JS}\right)^2}} \;
    \nu_{i}\left(\sqrt{q^2-\left(\frac{\omega}{JS}\right)^2} \right).
    \label{eq:AFM_dos_from_adjacency}
\end{equation}
The resulting antiferromagnetic local magnon density of states is shown in Fig.~\ref{fig:AFM_LDOS}, including the two bipartite $\{6,3\}$ and $\{4,4\}$ Euclidean lattices.
For a Heisenberg antiferromagnet in flat space, the long-wavelength dispersion is linear, $\omega_{\vb*{k}}\sim |\vb*{k}|$~\cite{anderson1952,kubo1952}.
Thus in two spatial dimensions, the bulk spectrum is gapless
\begin{equation}
    \rho^{\rm AFM}_{\rm Euc.}(\omega)\sim \omega,
    \qquad
    \omega\to 0^+ .
    \label{eq:AFM_Euclidean_DOS_scaling}
\end{equation}
The hyperbolic spectra in Fig.~\ref{fig:AFM_LDOS} are qualitatively different. 
Although the ordered state still breaks the global spin-rotation symmetry and therefore has isolated Goldstone modes, these modes carry only $O(1/N)$ local spectral weight. 
They therefore do not become a finite-density bulk continuum extending down to zero frequency. 
Instead, the thermodynamic bulk density of states is separated from zero frequency by a finite spectral gap $\Delta$, leading to a square-root lower band edge singularity,
\begin{equation}
    \rho^{\rm AFM}_{\rm hyp.}(\omega)\sim \sqrt{\omega-\Delta},
    \qquad
    \omega\to \Delta^+ .
    \label{eq:AFM_hyperbolic_DOS_scaling}
\end{equation}
This edge behavior is markedly different from the Euclidean case, where the two-dimensional Goldstone continuum gives $\rho(\omega)\sim\omega$ as $\omega\to 0^+$. 
It is also not the behavior obtained by simply adding a relativistic mass gap to a two-dimensional continuum dispersion, $\omega_{\vb*{k}}\sim\sqrt{c^2|\vb*{k}|^2+\Delta^2}$, which would give a constant density of states at the lower edge. 
A square-root onset is instead the band edge behavior associated with a massive relativistic dispersion in \emph{three} spatial dimensions.
In the present problem, it follows from the adjacency matrix spectrum of the hyperbolic lattice itself.
As shown in App.~\ref{subsec:Band-edge_scaling}, density of states moment asymptotics on hyperbolic lattices generically imply a square-root adjacency band edge, $\nu_i(\lambda)\sim |\lambda-\lambda_+|^{1/2}$, which is inherited by the magnon density of states under the smooth change of variables from adjacency eigenvalue to magnon frequency.
The oscillatory structure visible in the density of states for some finite-$p$ lattices originates from corresponding structure in the bulk adjacency density of states.
For example, the oscillations in the $\{8,3\}$ ferromagnetic density of states, shown in Fig.~\ref{fig:FM_DOS}, are mapped through the nonlinear transformation in Eq.~\eqref{eq:AFM_dos_from_adjacency} into the more pronounced oscillations of the antiferromagnetic magnon density of states. These features are stable with increasing cluster size and continued-fraction depth.

The density of states for the ferromagnet ($J<0$) is obtained even more directly. 
For the ferromagnet, the magnon energy is $\omega=|J|S(q-\lambda)$, so that
\begin{equation}
    \rho_i^{\rm FM}(\omega)
    =
    \frac{1}{|J|S}
    \nu_i\!\left(q-\frac{\omega}{|J|S}\right).
    \label{eq:FM_dos_from_adjacency}
\end{equation}
The ferromagnetic spectra are shown in App.~\ref{appendix:LDOS}, Fig.~\ref{fig:FM_DOS}. 
They coincide with the local density of states of the corresponding hyperbolic tight-binding problem up to this shift and rescaling of energy, consistent with previous studies of tight-binding spectra and hyperbolic band theory~\cite{mosseri2023,lux2024,lenggenhager2023,huang2025}.
As in the antiferromagnetic case, the hyperbolic ferromagnetic bulk spectrum has a finite lower band edge with a square-root onset $\rho^{\rm FM}_{\rm hyp.}(\omega)\sim \sqrt{\omega-\Delta}$. 
This again contrasts with the two-dimensional Euclidean ferromagnet, where $\omega_{\vb*{k}}\sim |\vb*{k}|^2$ and hence
\begin{equation}
    \rho^{\rm FM}_{\rm Euc.}(\omega)\sim {\rm const.},
    \qquad
    \omega\to 0^+ .
    \label{eq:FM_Euclidean_DOS_scaling}
\end{equation}

\begin{figure}[t!]
    \centering
    \begin{overpic}[width=\columnwidth]{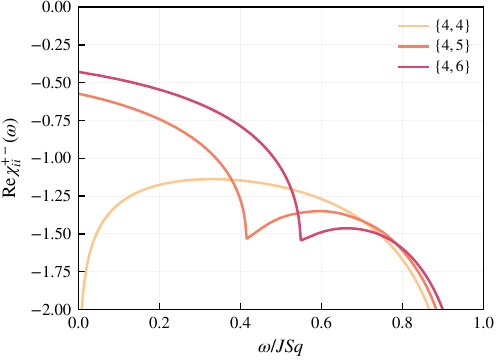}
    \end{overpic}
    \caption{
    Real part of the local transverse susceptibility for the bipartite Heisenberg antiferromagnet, as defined in Eq.~\eqref{eq:real_chi_from_dos}.
    The Euclidean square lattice $\{4,4\}$ shows a logarithmic infrared divergence arising from the gapless Goldstone continuum. 
    In contrast, the hyperbolic $\{4,5\}$ and $\{4,6\}$ lattices remain finite as $\omega \to 0^+$ because the thermodynamic bulk local density of states has a finite gap.
    }
    \label{fig:AFM_ReChi_omega}
\end{figure}

The distinction between isolated Goldstone modes and the bulk continuum is central to the bulk physics of the hyperbolic lattice model. 
For the ferromagnet, the uniform adjacency eigenvector at $\lambda=q$ is mapped by Eq.~\eqref{eq:FM_solution} to $\omega=0$. 
For the bipartite antiferromagnet, the two extremal adjacency eigenvalues $\lambda=\pm q$ are both mapped by Eq.~\eqref{eq:AFM_solution} to $\omega=0$, corresponding to the uniform and staggered global $\mathrm{SO}(3)$ rotations in the locally rotated frame. 
These modes carry local weight of order $1/N$ at any fixed bulk site and therefore disappear from the bulk local density of states in the thermodynamic limit.
The infrared structure of local observables is instead controlled by the lower edge of the bulk continuum.
This separation between the global zero mode(s) and gapped local response is closely analogous to the absence of Goldstone bosons in local correlations on the Bethe lattice~\cite{laumann2009}. 
Related bulk-gap phenomena have also been discussed in the context of vector-field excitations on hyperbolic lattices~\cite{sela2025}.

The local (retarded) transverse response provides a direct way to probe the infrared
structure of the magnon excitations. 
We define
\begin{equation}
    \chi_{ii}^{+-}(\omega)
    \equiv
    -i \int_0^\infty \dd t\,
    e^{i(\omega+i0^+)t}
    \expval{\comm{S_i^+(t)}{S_i^-(0)}} .
    \label{eq:local_transverse_susceptibility}
\end{equation}
Here the index $i$ denotes a fixed bulk site.
Within linear spin-wave theory, the positive-frequency spectral weight (dissipative part) of this response is proportional to the local magnon density of states,
\begin{equation}
    -\frac{1}{\pi}\Im \chi_{ii}^{+-}(\omega)
    =
    2 u^2(\omega) \, \rho_i(\omega),
    \qquad \omega>0 ,
    \label{eq:im_chi_dos_relation}
\end{equation}
where $u(\omega)$ is the Bogoliubov coherence factor defined in Eqs.~\eqref{eq:FM_solution} and~\eqref{eq:AFM_solution}.
The real part is then obtained from the same spectral information using the Kramers--Kronig relation~\cite{altland2010},
\begin{equation}
    \Re \chi_{ii}^{+-}(\omega)
    =
    2\,\mathcal{P}\int_0^\infty \dd\omega'\,\rho_i(\omega')
    \left[
        \frac{u^2(\omega')}{\omega-\omega'}
        -
        \frac{v^2(\omega')}{\omega+\omega'}
    \right],
    \label{eq:real_chi_from_dos}
\end{equation}
where $\mathcal{P}$ denotes the Cauchy principal value.

Equation~\eqref{eq:real_chi_from_dos} shows why the Euclidean and hyperbolic
cases differ so sharply.
For both the two-dimensional Euclidean ferromagnet and antiferromagnet, the weighted spectral density entering Eq.~\eqref{eq:real_chi_from_dos} approaches a constant at low frequency,
\begin{equation}
\rho_i(\omega)\left[ u^2(\omega)+v^2(\omega)\right] \sim{\rm const.}, \qquad \omega \to 0^+,
\end{equation}
producing the logarithmic singularity
\begin{equation}
    \Re \chi_{ii}^{+-}(\omega)
    \sim
    -\ln\frac{\omega_*}{|\omega|},
    \qquad
    \omega \to 0^+,
    \label{eq:log_chi}
\end{equation}
where $\omega_*$ is a nonuniversal ultraviolet frequency scale of order the magnon bandwidth.
This logarithmic infrared singularity is the local spectral signature of the Euclidean Goldstone continuum.

In contrast, the hyperbolic lattices have a finite bulk infrared cutoff due to the finite spectral gap.
As a result, the static local response $\Re\chi_{ii}^{+-}(0)$ remains finite even though the ordered state still has isolated zero-energy Goldstone modes associated with global $\mathrm{SO}(3)$ spin rotations. 
This coexistence of symmetry-required zero modes with a finite bulk local transverse response is a distinctive feature of the hyperbolic problem --- the Goldstone modes are collective modes with vanishing local spectral weight, while the thermodynamic bulk local spectrum is gapped. 
This behavior is illustrated in Fig.~\ref{fig:AFM_ReChi_omega}, where the Euclidean $\{4,4\}$ antiferromagnet exhibits the expected logarithmic infrared divergence, while the hyperbolic $\{4,5\}$ and $\{4,6\}$ lattices saturate to finite values as $\omega\to0$. 
The finite-frequency structure of the hyperbolic lattice curves is not associated with an additional low-energy mode, but reflects the Kramers--Kronig response to the gapped bulk magnon continuum, whose band-edge onset and higher-frequency spectral weight both enter the principal-value integral in Eq.~\eqref{eq:real_chi_from_dos}.
Since $\chi^{+-}_{ii}$ is a local bulk response, exponential volume growth does not enter through an explicit sum over all lattice sites, but indirectly through its effect on the bulk magnon spectrum and local density of states.

The analysis above shows that the key distinction between Euclidean and hyperbolic spin waves is the separation between the isolated global zero mode and the lower edge of the bulk continuum. 
The size of this bulk gap controls both the onset of the density of states and the saturation of the local transverse response. 
We therefore next quantify how the curvature-induced infrared cutoff appears in the spin-wave spectrum by studying the dependence of the bulk magnon gap on the intrinsic curvature scale $|\kappa|a^2$ of the underlying $\{p,q\}$ lattice [cf.~Eq.~\eqref{eq:pq_curvature_relation}].

\subsection{Magnon gap}
\label{subsec:magnon_gap}

\begin{figure}[t!]
    \centering
    \begin{overpic}[width=\columnwidth]{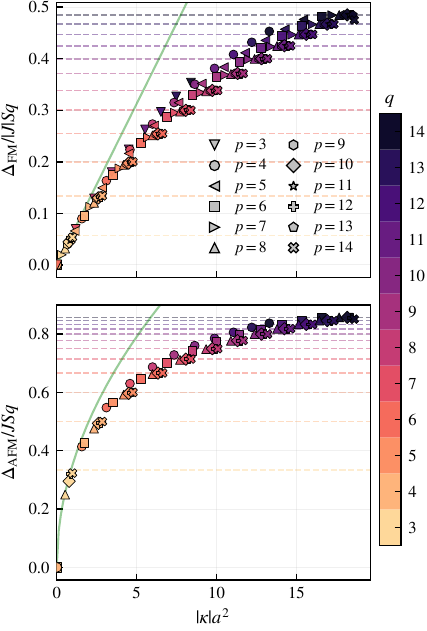}
    \end{overpic}
    \caption{
    Bulk magnon gap at $O(S)$ as a function of the intrinsic curvature scale $|\kappa|a^2$ for the hyperbolic $\{p,q\}$ lattices.
    The upper panel shows the ferromagnet, while the lower panel shows the bipartite antiferromagnet.
    Symbols distinguish $p$, colors distinguish $q$, and dashed horizontal lines denote the corresponding $q$-regular Bethe lattice values discussed in App.~\ref{app:BetheLattice}.
    The green lines represent the continuum limits of Eqs.~\eqref{eq:AFM_continuum_gap_main} and~\eqref{eq:FM_continuum_gap_main}.
    }
    \label{fig:gap}
\end{figure}

We now quantify the bulk magnon gap associated with the infrared cutoff identified in the bulk density of states.
The results are shown in Fig.~\ref{fig:gap} for both the ferromagnet and the bipartite antiferromagnet.
In the Euclidean cases, which occur at $|\kappa|a^2=0$, the bulk magnon spectrum is gapless.
This is the expected situation in flat two-dimensional spin systems, where the Goldstone mode is the endpoint of a continuum of long-wavelength spin waves.

Conversely, for every hyperbolic $\{p,q\}$ lattice shown in Fig.~\ref{fig:gap}, the thermodynamic bulk spectrum has a finite gap.
At fixed coordination $q$, the gap generally becomes larger as $p$ is increased. 
The gap increases \emph{approximately} monotonically as the intrinsic curvature scale $|\kappa|a^2$ is increased, although small variations are visible in the numerical data.
This same overall trend holds for both the ferromagnetic and antiferromagnetic models, suggesting that the presence of a gap is not tied to the particular microscopic model, but to the underlying hyperbolic graph geometry.

Crucially, this gap does not arise from spin-rotation anisotropy.
The Hamiltonian in Eq.~\eqref{eq:HeisenbergModel} remains fully $\mathrm{SO}(3)$-invariant, and the ordered state still has zero-energy collective rotations. 
The gap shown in Fig.~\ref{fig:gap} is instead the separation between these isolated Goldstone modes and the lower edge of the thermodynamic bulk magnon continuum. 
Because the local weight of each isolated symmetry mode scales as $1/N$, these modes disappear from the bulk local density of states as $N\to\infty$. 
The finite gap in Fig.~\ref{fig:gap} should therefore be understood as a bulk local spectral gap, not as a gap produced by explicit breaking of spin-rotation symmetry.

Furthermore, this gap is \emph{not} a finite-size effect. 
Our calculation probes the bulk local density of states, explicitly excluding the boundary modes of a finite open hyperbolic graph.
This distinction is essential because finite hyperbolic clusters have an extensive number of boundary vertices~\cite{zhang2022}, and boundary-localized modes can obscure the bulk infrared structure~\cite{chen2023a,sela2025}.
The gap plotted in Fig.~\ref{fig:gap} instead characterizes the bulk properties of the lattice in the thermodynamic limit.

The geometric origin of the gap is analogous to the spectral gap of kinetic-energy operators in negatively curved space. 
For example, the Laplace--Beltrami operator on the hyperbolic plane has a continuum of gapped eigenvalues, with the gap set by the Gaussian curvature~\cite{mckean1970,chavel1984,camporesi1994,sarnak2003}.
On a lattice, the corresponding effect appears as a separation between the isolated zero-energy mode and the lower edge of the bulk magnon continuum.
This is the same mechanism that underlies the absence of dispersive Goldstone bosons on the Bethe lattice~\cite{laumann2009}, a connection made explicit in App.~\ref{app:BetheLattice}.

Finally, the $q$-regular Bethe lattice provides the limiting case of this curvature-induced gap.
For fixed $q$, increasing $p$ moves the $\{p,q\}$ lattice toward the $\{\infty,q\}$ Bethe lattice limit, and the gaps in Fig.~\ref{fig:gap} approach the corresponding Bethe lattice values.
These values provide upper bounds on the bulk magnon gaps of the $q$-regular $\{p,q\}$ lattices,
\begin{equation}
    \Delta_{\rm FM}(p,q) \leq \Delta_{\rm FM}^{\rm Bethe}(q),
    \qquad
    \Delta_{\rm AFM}(p,q) \leq \Delta_{\rm AFM}^{\rm Bethe}(q),
    \label{eq:bethe_gap_bound}
\end{equation}
as derived in App.~\ref{app:BetheLattice}, where the explicit expressions are also given.
Having established the curvature-induced spectral gap, we now examine its consequences for the quantum ground state.

\subsection{Zero-point fluctuations}
\label{subsec:ZeroPointFluctuations}

We now turn from the magnon spectrum itself to the quantum renormalization of the ground state. 
For the ferromagnet, the fully polarized ground state is an exact eigenstate of the Hamiltonian, so there are no zero-point quantum fluctuations.
We therefore restrict our discussion at $T=0$ to the bipartite antiferromagnet.
Although the N\'eel state minimizes the classical energy on bipartite $\{p,q\}$ lattices, it is not an exact eigenstate of the quantum Hamiltonian. 
The anomalous terms in Eq.~\eqref{eq:H2_site_basis} generate a Bogoliubov vacuum with finite zero-point magnon occupation, producing the familiar $T=0$ quantum corrections to the ground-state energy and staggered magnetization~\cite{anderson1952,manousakis1991,castilla1991,igarashi1992,gochev1993,igarashi2005}. 
Here we examine how these corrections depend on the negative curvature of the underlying hyperbolic plane.

\begin{figure}[tbp]
    \centering
    \includegraphics[width=\columnwidth]{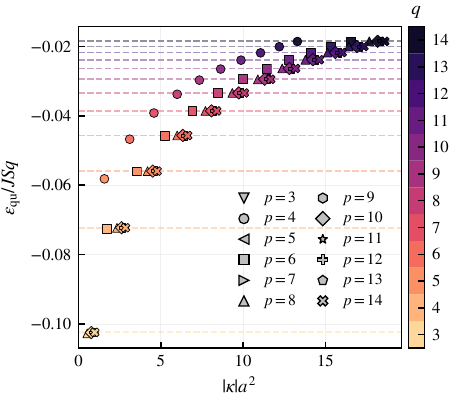}
    \caption{
    Quantum zero-point correction $\varepsilon_{\rm qu}$ to the antiferromagnetic ground-state energy per spin, as defined in Eq.~\eqref{eq:AFM_zero_point_energy}.
    Increasing negative curvature reduces the magnitude of the quantum energy lowering, driving the system toward more mean-field-like behavior.
    The exact Bethe lattice solution is depicted by the dashed lines, giving the limiting value at fixed $q$.
    }
    \label{fig:AFM_ZeroPoint}
\end{figure}

From the diagonalized Bogoliubov Hamiltonian in Eq.~\eqref{eq:diagonalized_H}, the ground-state energy can be written as
\begin{equation}
    \expval{H} =
    \left(1+\frac{1}{S}\right)\mathcal{E}_{\rm cl} + \frac{1}{2}\sum_{n}\omega_n + O(S^0),
    \label{eq:expec_H}
\end{equation}
with the expectation value taken at $T=0$. 
The $O(S^2)$ part of Eq.~\eqref{eq:expec_H} is the classical contribution to the ground-state energy $\mathcal{E}_{\rm cl} = -JS^2Nq/2$.
The $O(S)$ part of Eq.~\eqref{eq:expec_H} gives the leading $1/S$ quantum correction to the ground-state energy.
We define the zero-point energy per spin as 
\begin{equation}
    \varepsilon_{\rm qu}
    \equiv
    \frac{\mathcal{E}_{\rm cl}}{NS} +
    \frac{1}{2}
    \int_0^\infty \dd\omega\,
    \omega \;
    \rho^{\rm AFM}(\omega).
    \label{eq:AFM_zero_point_energy}
\end{equation}
This correction is negative because quantum fluctuations lower the energy relative to the classical N\'eel state. 
Figure~\ref{fig:AFM_ZeroPoint} shows that the magnitude of this quantum correction generally decreases as the curvature $|\kappa|a^2$ increases.
Within linear spin-wave theory, increasing curvature suppresses the zero-point correction, driving the system toward the classical limit.
This follows from the curvature-induced suppression of low-energy spectral weight, which shifts the first moment of the magnon density of states in Eq.~\eqref{eq:AFM_zero_point_energy} to higher energies.

\begin{figure}[tbp]
    \centering
    \includegraphics[width=\columnwidth]{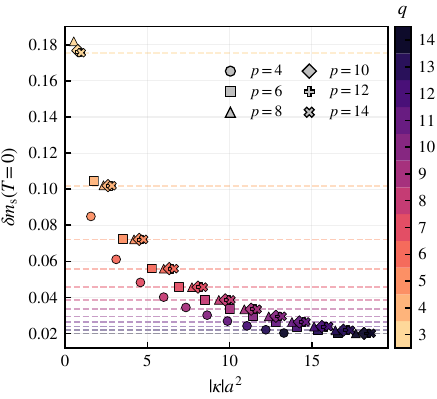}
    \caption{
    Zero-temperature correction $\delta m_s(0)$ to the staggered magnetization per spin for the bipartite Heisenberg antiferromagnet, as defined in Eq.~\eqref{eq:AFM_zero_point_magnetization}. 
    The correction decreases with increasing negative curvature, indicating suppressed zero-point magnon occupation.
    }
    \label{fig:AFM_magnetization_T0}
\end{figure}

Quantum zero-point fluctuations also reduce the staggered magnetization per spin, obtained from Eq.~\eqref{eq:HP_representation} as
\begin{equation}
    \expval{S_i \cdot \vu*{e}_{i,0}}=S-\delta m_s(0),
\end{equation}
where
\begin{equation}
    \delta m_s(0)
    \equiv
    \expval{a_i^\dagger a_i^{}}
    =
    \int_0^\infty \dd\omega\,
    \rho^{\rm AFM}(\omega) 
    \; v^2(\omega),
    \label{eq:AFM_zero_point_magnetization}
\end{equation}
with the coherence factor $v(\omega)$ defined in Eq.~\eqref{eq:AFM_solution}.
Equation~\eqref{eq:AFM_zero_point_magnetization} is infrared-sensitive, since the coherence factor diverges as $v^2(\omega)\sim 1/\omega$ for small $\omega$.
As shown in Fig.~\ref{fig:AFM_magnetization_T0}, more negative curvature suppresses the zero-point magnon occupation and therefore increases the ordered moment.

At fixed coordination $q$, both the quantum zero-point energy $\varepsilon_{\rm qu}$ and correction to the staggered magnetization $\delta m_s(0)$ approach their Bethe lattice values as the curvature is increased. 
The Bethe lattice therefore provides the limiting bound for the family of $\{p,q\}$ lattices: it gives the least negative zero-point energy correction,
\begin{equation}
    \varepsilon_{\rm qu}^{\{p,q\}}
    \leq
    \varepsilon_{\rm qu}^{\rm Bethe}(q)
    <0,
\end{equation}
and the smallest order parameter correction,
\begin{equation}
    \delta m_s^{\{p,q\}}(0)
    \geq
    \delta m_s^{\rm Bethe}(q)
    >
    0.
\end{equation}
Thus the general effect of increasing negative curvature is to suppress quantum fluctuations and drive the antiferromagnet toward a more mean-field-like ordered state. 
The analytic Bethe lattice expressions are derived in App.~\ref{app:BetheLattice}.

The zero-temperature results show that negative curvature suppresses quantum fluctuations of the ordered moment. 
Since the same low-energy spin-wave spectrum controls the thermal depletion, this raises the question of whether the curvature-induced infrared cutoff also regularizes the finite-temperature spin-wave population that destroys long-range order in the Euclidean two-dimensional Heisenberg model.

\subsection{Finite-temperature transition}
\label{subsec:finite_temperature_transition}

We now use the bulk magnon spectrum to estimate the magnetic transition temperature.  
This provides one of the central results of the paper. 
In contrast to the conventional two-dimensional Euclidean Heisenberg model, the hyperbolic lattices support a finite spin-wave estimate of the transition temperature.
The essential mechanism is that negative curvature separates the isolated global symmetry mode from the bulk magnon continuum, producing a finite infrared cutoff in the local bulk density of states.

Within linear spin-wave theory, the ordered moment is generally reduced by zero-point and thermal fluctuations by an amount
\begin{align}
    &\delta m(T)
    \equiv
    S - \expval{S_i \cdot \vu*{e}_{i,0}} \nonumber \\
    &\quad=
    \int_0^\infty \dd\omega\,
    \rho(\omega)
    \left[
        v^2(\omega)
        +
        \left(
            u^2(\omega)+v^2(\omega)
        \right)
        n_{\rm B}\left( \frac{\omega}{T} \right)
    \right],
    \label{eq:thermal_magnetization}
\end{align}
where $n_{\rm B}(z) \equiv 1/(e^{z}-1)$ is the Bose distribution function and we take $k_{\rm B}=1$.
We define the spin-wave estimate of the critical temperature $T_{\rm c}^*$ at $O(S)$ by the condition that spin-wave fluctuations completely exhaust the ordered moment,
\begin{equation}
    S
    =
    \delta m(T_{\rm c}^*).
    \label{eq:Tc_criterion}
\end{equation}
Note that the solution to Eq.~\eqref{eq:Tc_criterion} should not be interpreted as a controlled calculation of the true critical temperature. 
As the temperature approaches the true critical temperature from below, $T \to T_{\rm c}^-$, magnon--magnon interactions at higher order in $1/S$~\cite{dyson1956a,oguchi1960} and critical fluctuations~\cite{chaikin1995} render a single-magnon description unreliable. 
Equation~\eqref{eq:Tc_criterion} therefore provides an estimate of the critical temperature by extrapolating from the low-temperature spin-wave theory.

\begin{figure}[tbp]
    \centering
    \begin{overpic}[width=\columnwidth]{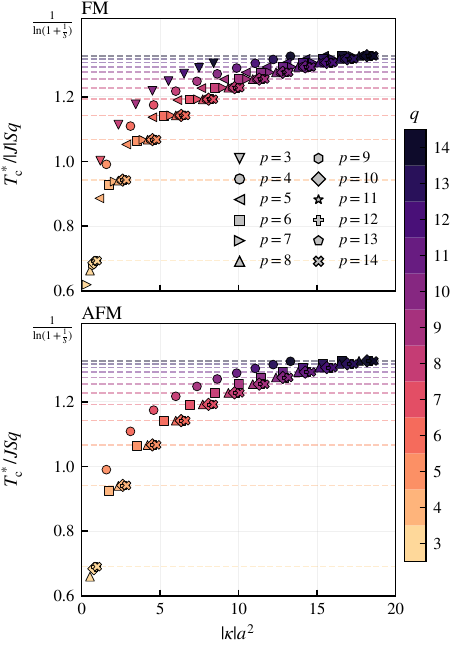}
    \end{overpic}
    \caption{
    Spin-wave estimate of the critical temperature $T_{\rm c}^*$ for the ferromagnetic (top panel) and bipartite antiferromagnetic (bottom panel) Heisenberg models on hyperbolic $\{p,q\}$ lattices, computed for $S=1$. 
    The estimate is obtained from the thermal spin-wave depletion criterion in Eq.~\eqref{eq:Tc_criterion}. 
    Dashed lines denote the corresponding $q$-regular Bethe lattice values obtained in the formal limit $p\to\infty$, which bound the hyperbolic-lattice estimates according to $T_{\rm c}^* < |J|Sq/\ln(1+1/S)$, as derived in App.~\ref{app:BetheLattice}. 
    The increase of $T_{\rm c}^*$ with increasing $|\kappa|a^2$ reflects the curvature-induced suppression of low-energy spin-wave fluctuations.
    }
    \label{fig:Tc}
\end{figure}

The solutions to Eq.~\eqref{eq:Tc_criterion} are shown in Fig.~\ref{fig:Tc} for both the ferromagnetic and bipartite antiferromagnetic models. 
In both cases, increasing the curvature scale $|\kappa|a^2$ tends to increase $T_{\rm c}^*$, indicating that negative curvature stabilizes magnetic order by suppressing the infrared spin-wave fluctuations that destabilize ordering in flat two-dimensional space. 
For fixed coordination number $q$, the corresponding $q$-regular Bethe lattice, obtained formally in the limit $p\to\infty$, provides an upper bound on the estimated transition temperature. 

The distinction between Euclidean and hyperbolic lattices is apparent from the infrared structure of Eq.~\eqref{eq:thermal_magnetization}. 
For the two-dimensional Euclidean ferromagnet $(J<0)$, the quadratic Goldstone dispersion leads to a constant density of states at low energy $\rho^{\rm FM}(\omega)\to \mathrm{const.}$ as $\omega \to 0^+$. 
For the two-dimensional Euclidean antiferromagnet $(J>0)$, $\rho^{\rm AFM}(\omega)\sim \omega$ at low energy, but $u^2(\omega)+v^2(\omega)\sim 1/\omega$, so the product entering the thermal part of Eq.~\eqref{eq:thermal_magnetization} is constant. 
In either case,
\begin{equation}
    \rho(\omega)
    \left[
    u^2(\omega)+v^2(\omega)
    \right]
    \to
    \mathcal{C}_0,
    \qquad
    \omega\to 0^+,
\end{equation}
for some nonuniversal positive constant $\mathcal{C}_0$.
At any finite temperature, the ordered moment thus becomes infrared divergent
\begin{equation}
    \delta m(T)
    \sim
    \delta m(0)
    +
    \mathcal{C}_0 T
    \int_0^{\omega_*} \frac{\dd\omega}{\omega}.
    \label{eq:euclidean_IR_divergence}
\end{equation}
This logarithmic divergence is a manifestation of the Mermin--Wagner--Hohenberg theorem, which forbids spontaneous breaking of a continuous symmetry at finite temperature in one- and two-dimensional Euclidean systems with short-ranged interactions~\cite{mermin1966,hohenberg1967}.
The logarithmically divergent nature of $\delta m(T)$ for Euclidean lattices implies there is no finite-temperature solution to Eq.~\eqref{eq:Tc_criterion}.

For the hyperbolic $\{p,q\}$ lattices, the thermodynamic bulk spectrum is gapped, beginning at a finite lower band edge $\Delta$. 
The isolated Goldstone modes appear only through the explicit $O(1/N)$ delta-function terms in the finite-size density of states and do not contribute finite weight to the local bulk density of states in the thermodynamic limit.
As derived in App.~\ref{appendix:LDOS}, the band edge has a square-root singularity, and thus
\begin{equation}
    \rho(\omega)
    \left[
        u^2(\omega)+v^2(\omega)
    \right]
    \sim
    \mathcal{C}
    \sqrt{\omega-\Delta},
    \qquad
    \omega\to \Delta^+ ,
\end{equation}
for some nonuniversal positive constant $\mathcal{C}$.
Consequently, the infrared part of the ordered moment is
\begin{equation}
    \delta m(T)
    \sim
    \delta m(0)
    +
    \mathcal{C}\,
    \frac{\sqrt{\pi}}{2}\,
    T^{3/2} 
    e^{-\Delta/T}.
    \label{eq:activated_thermal_depletion}
\end{equation}
This form is reminiscent of the low-temperature magnetization of a \emph{three-dimensional} Heisenberg magnet with a weak anisotropy gap, where the square-root density of states gives a $T^{3/2}$ prefactor while the anisotropy produces the activated behavior $\sim e^{-\Delta/T}$. 
The distinction is that such a gap in a Euclidean Heisenberg magnet requires explicitly breaking the spin-rotation symmetry, whereas Eq.~\eqref{eq:activated_thermal_depletion} arises here in a \emph{two-dimensional} lattice while preserving the global $\mathrm{SO}(3)$ symmetry. 
The activated temperature dependence is instead generated by the curvature-induced separation between the isolated global symmetry mode and the bulk magnon continuum.

Thus far, the role of the bulk magnon gap has been diagnosed through local quantities. 
We now turn to the complementary spatial question of how transverse spin correlations behave at large distances in hyperbolic space.

\subsection{Transverse correlations}
\label{subsec:transverse_correlations}

Section~\ref{subsec:magnon_gap} established that the magnon spectrum of the bipartite Heisenberg antiferromagnet on a hyperbolic lattice has isolated global symmetry modes separated from the thermodynamic bulk magnon continuum by a finite gap. 
We now examine the corresponding real-space consequence in the equal-time transverse spin-spin correlations. 
This provides a direct diagnostic of the distinction between the hyperbolic antiferromagnet and the conventional Euclidean two-dimensional antiferromagnet: in flat space, gapless Goldstone modes produce an algebraic transverse tail, whereas on the hyperbolic lattice the local bulk spectrum is gapped and the transverse correlations are short ranged.

We define the equal-time transverse correlation function by
\begin{equation}
    C_\perp(r_{ij})
    \equiv
    \frac{1}{2S}
    \expval{S_i^+ S_j^- + S_i^- S_j^+} ,
    \label{eq:AFM_transverse_correlation}
\end{equation}
where $r_{ij}$ denotes the graph distance between sites $i$ and $j$.
The transverse component contains the leading spatially dependent correlations within linear spin-wave theory.  
In contrast, the leading longitudinal contribution is the disconnected ordered moment, $m_s^2(0)$, while nontrivial connected longitudinal correlations first arise at $O(S^0)$~\cite{canali1993}. 
We therefore restrict attention to the transverse sector, which also exhibits most clearly the key differences between spontaneous symmetry breaking on Euclidean vs hyperbolic lattices.

For the bipartite antiferromagnet, Eq.~\eqref{eq:AFM_transverse_correlation} alternates in sign with sublattice parity as it is defined in the global frame.
Since we are interested in the long-distance envelope of the correlations, we consider the magnitude $|C_\perp(r)|$.
For the finite graphs used in this work, we estimate the corresponding bulk value by averaging $|C_\perp(r_{ij})|$ over pairs of bulk sites at fixed graph distance to reduce boundary effects and improve numerical stability.
When evaluating Eq.~\eqref{eq:AFM_transverse_correlation}, the isolated global zero modes are excluded. 
This is physically equivalent to fixing the collective orientation of the symmetry-broken N\'eel state before computing connected local correlations (say, the $\vu*{z}$ axis, with $S_j^\pm=S_j^x\pm iS_j^y$). 
The resulting correlator is governed by the bulk magnon continuum rather than by the finite-size collective coordinate.

\begin{figure}[tbp]
    \centering
    \begin{overpic}[width=\columnwidth]{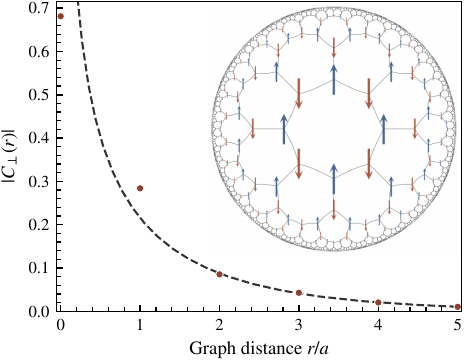}
    \end{overpic}
    \caption{
    Magnitude of the equal-time transverse correlation function defined in Eq.~\eqref{eq:AFM_transverse_correlation} for the antiferromagnetic Heisenberg model on the $\{8,3\}$ lattice at zero temperature.
    The correlator is evaluated between bulk sites separated by graph distance $r/a$ on a large finite cluster, the plotted values are obtained by averaging over all bulk pairs at fixed distance. 
    The dashed curve is a fit to the asymptotic form in Eq.~\eqref{eq:AFM_hyperbolic_transverse_fit}, giving $A_\perp=0.38$ and $\xi_\perp/a=1.8$. 
    The finite transverse correlation length shows that local transverse correlations are short ranged, in contrast to the $1/r$ Goldstone tail of the Euclidean two-dimensional antiferromagnet.
    }
    \label{fig:AFM_transverse_correlation}
\end{figure}

The resulting transverse correlation function for the $\{8,3\}$ lattice is shown in Fig.~\ref{fig:AFM_transverse_correlation}.
We find the large-distance behavior is well described by the function
\begin{equation}
    |C_\perp(r)|
    \sim
    A_\perp
    \frac{e^{-r/\xi_\perp}}{\sqrt{r/a}} .
    \label{eq:AFM_hyperbolic_transverse_fit}
\end{equation}
Fitting over the accessible bulk range gives $A_\perp = 0.38$ and a finite \emph{transverse} correlation length $\xi_\perp/a = 1.8$.
The asymptotic form of Eq.~\eqref{eq:AFM_hyperbolic_transverse_fit} matches the exact Bethe-lattice result derived in App.~\ref{app:BetheLattice}.
On a $q$-regular Bethe lattice, the same large-distance form is obtained with $\xi_{\perp}=a/\ln(q-1)$. 
A related distinction between massive local correlations and singular global response appears in the Bethe lattice spherical model of Ref.~\cite{laumann2009}.
Note that the correlation length considered here is a zero-temperature quantity. 
Its temperature dependence, particularly near the thermal transition where linear spin-wave theory becomes unreliable, is beyond the scope of the present work.

The form of Eq.~\eqref{eq:AFM_hyperbolic_transverse_fit} should be contrasted with the infinite transverse correlation length $\xi_\perp\rightarrow\infty$ in the conventional two-dimensional Euclidean antiferromagnet, as usually expected for the spontaneous breakdown of a continuous symmetry. 
Near the ordering wave vector, the spin-wave dispersion is linear, $\omega_{\vb*{k}}\sim |{\vb*{k}}|$, and the equal-time transverse correlator thus takes the asymptotic form 
\begin{equation}
    |C_\perp^{\rm Euc}(r)|
    =
    JqV\int \frac{\dd^2 k}{(2\pi)^2}
    \frac{e^{i{\vb*{k}}\cdot{\vb*{r}}}}{\omega_{\vb*{k}}}
    \sim
    \frac{\sqrt{2}\,V}{2\pi a} \frac{1}{r},
    \label{eq:Euclidean_transverse_goldstone_tail}
\end{equation}
where $V$ is the magnetic unit cell volume. 
This algebraic $1/r$ tail is the equal-time real-space signature of the continuum of Goldstone modes in a conventional two-dimensional antiferromagnet~\cite{canali1993,patashinskii1973}.

The hyperbolic antiferromagnet is therefore qualitatively distinct from its flat-space counterpart. 
Although the broken $\mathrm{SO}(3)$ symmetry produces isolated global zero modes, local transverse correlations are controlled by the gapped bulk magnon continuum. 
The result is a finite transverse correlation length rather than the $1/r$ Goldstone tail of the Euclidean two-dimensional antiferromagnet.

The coexistence of a finite transverse correlation length with a divergent bulk response is a characteristic feature of hyperbolic space.
Conversely, in Euclidean systems, a divergent susceptibility is normally tied to $\xi_\perp\to\infty$.
This distinction is tied to the fact that the number of sites at distance $r$ grows
exponentially. 

This point should be distinguished from the \emph{local} transverse
susceptibility $\chi_{ii}^{+-}$ discussed in Sec.~\ref{subsec:local_transverse_susceptibility}.
The local response probes the spectral weight at a fixed bulk site and is insensitive to the isolated Goldstone mode in the thermodynamic limit, since that mode has local weight of order $1/N$. 
By contrast, the equal-time structure factor probes spatially coherent fluctuations of the ordered moment and therefore involves a sum over correlations throughout the lattice. 
This is the sense in which a volume-divergent structure factor indicative of global long-range order can coexist with finite local response and finite transverse correlation length.
The same distinction underlies the Bethe lattice spherical model of Ref.~\cite{laumann2009}, where local correlation functions remain massive even in the symmetry-broken phase, while global quantities retain the singular response associated with broken $\mathrm{SO}(3)$ symmetry.

We now make this statement explicit, building on the same logic as Ref.~\cite{morita1975}. 
Let $\mathcal{Q}_n$ denote the number of sites at graph distance $r=na$ from a reference bulk site. 
For a hyperbolic lattice,
\begin{equation}
    \mathcal{Q}_n \sim \lambda^n
    =
    e^{n \ln \lambda}, \qquad n\gg 1,
    \label{eq:exponential_growth}
\end{equation}
where $\ln \lambda$ is the graph-volume entropy. 
This exponential growth is the discrete counterpart of the exponential area growth of geodesic spheres in the hyperbolic plane. 
For regular hyperbolic lattices, the growth rate $\lambda_{p,q}$ can be obtained from the corresponding shell-counting recurrence relation or generating function~\cite{floyd1994,bartholdi2002,chen2023a,mosseri2023}; we review this calculation in App.~\ref{app:ShellGrowth}.

Because the transverse correlator of a bipartite antiferromagnet alternates with sublattice parity, the shell average relevant to the N\'eel order parameter is the staggered correlation, with respect to a bulk reference site $i$,
\begin{equation}
    \overline{C}_{\perp}(na)
    =
    \frac{1}{\mathcal{Q}_n}
    \sum_{j}
    \delta_{na,r_{ij}}
    \eta_i\eta_j\, C_\perp(r_{ij}),
\end{equation}
where $\eta_i=\pm1$ alternates on the two sublattices and $C_{\perp}(r_{ij})$ is defined in Eq.~\eqref{eq:AFM_transverse_correlation}.
The staggered equal-time structure factor can then be written as
\begin{equation}
    \mathcal{S}_{\rm bulk}^{\perp}
    =
    \sum_{n=0}^{\infty}
    \mathcal{Q}_n\,\overline{C}_{\perp}(na).
    \label{eq:shell_structure_factor}
\end{equation}
Using Eqs.~\eqref{eq:AFM_hyperbolic_transverse_fit} and~\eqref{eq:exponential_growth}, the large-$n$
contribution of a shell scales as
\begin{equation}
    \mathcal{Q}_n\,\overline{C}_{\perp}(na)
    \sim
    \frac{1}{\sqrt{n}}\,
    \exp\!\left[
        n\left(\ln\lambda-\frac{a}{\xi_\perp}\right)
    \right].
    \label{eq:shell_weight}
\end{equation}
The shell sum therefore diverges whenever
\begin{equation}
    \xi_\perp \geq \frac{a}{\ln\lambda}.
    \label{eq:hyperbolic_susceptibility_criterion}
\end{equation}
At the marginal value $\xi_\perp=a/\ln\lambda$, the exponential decay of the correlator is exactly balanced by the exponential growth of the shell volume, leaving a residual $n^{-1/2}$ shell contribution. 
The sum in Eq.~\eqref{eq:shell_structure_factor} then still diverges.
Thus the relevant criterion for a divergent structure factor on a hyperbolic lattice is not $\xi_\perp=\infty$, but rather whether $\xi_\perp/a$ exceeds the inverse graph-volume entropy.

For the $\{8,3\}$ lattice,
\begin{equation}
    \lambda_{8,3}
    =
    \frac{
    1+\sqrt{13}+\sqrt{2(\sqrt{13}-1)}
    }{4},
    \quad
    \frac{1}{\ln\lambda_{8,3}}
    \simeq 1.8398 .
    \label{eq:8_3_volume_entropy}
\end{equation}
This value is very close to the fitted transverse correlation length $\xi_\perp/a=1.8$ obtained in Fig.~\ref{fig:AFM_transverse_correlation}. 
Together with the exact Bethe lattice result $\xi_\perp=a/\ln(q-1)$ and the continuum result
$\xi_\perp=1/\sqrt{|\kappa|}$ derived in App.~\ref{app:AFM_continuum}, these results suggest a simple geometric conjecture. 
Namely, for the bipartite hyperbolic $\{p,q\}$ lattices considered here, the asymptotic transverse correlation length is set by the inverse graph-volume entropy,
\begin{equation}
    \xi_\perp
    \stackrel{!}{=}
    \frac{a}{\ln\lambda_{p,q}} .
    \label{eq:correlation_length_entropy_conjecture}
\end{equation}
Equivalently, the shell-growth criterion in Eq.~\eqref{eq:hyperbolic_susceptibility_criterion} is saturated. 
Our conjecture is exactly satisfied in both the Bethe-lattice and continuum limits, where the magnon gap and the volume entropy are controlled by the same curvature scale. 
Physically, this is natural, as the same local connectivity that drives exponential shell growth also lifts the bulk magnon spectrum away from zero frequency.
In this interpretation, local transverse correlations remain exponentially short ranged, but their shell-integrated staggered weight is marginal.
The exponential decay of the correlator is exactly balanced by the exponential growth of hyperbolic shells.

\subsection{Continuum limit}
\label{subsec:continuum}

The lattice spin-wave results discussed thus far admit a simple continuum interpretation in the low-curvature regime, where the lattice spacing $a$ is small compared to the radius of curvature $\ell \equiv 1/\sqrt{|\kappa|}$.
In this limit, the long-wavelength antiferromagnetic fluctuations are described by the $\mathrm{O}(3)$ nonlinear sigma model~\cite{chakravarty1988,ioffe1988} on the hyperbolic plane with (inverse) metric tensor $g^{ab}$,
\begin{equation}
    S_{\rm AFM}[\vb*{n}]
    =
    \frac{\rho_s}{4}
    \int \dd\tau \,\dd z\,\dd\bar z\sqrt{|g|}
    \left[
        \frac{\left(\partial_\tau \vb*{n}\right)^2}{c^2}
        +
        g^{ab}\partial_a\vb*{n}\cdot\partial_b\vb*{n}
    \right],
    \label{eq:AFM_continuum_action_main}
\end{equation}
where $|\vb*{n}|^2=1$ is the N\'eel order parameter, $c = JSqa/\sqrt{2}$ is the magnon velocity and 
\begin{equation}
    \rho_s = \frac{JS^2pq|\kappa|a^2}{4\pi\left[(p-2)(q-2)-4\right]}
\end{equation}
is the spin stiffness.
The derivation of Eq.~(\ref{eq:AFM_continuum_action_main}) from the microscopic coherent-state path integral is given in App.~\ref{app:AFM_continuum}.

Expanding Eq.~\eqref{eq:AFM_continuum_action_main} about a N\'eel ordered state gives two transverse Goldstone fields whose quadratic fluctuations are governed by the Laplace--Beltrami operator on the hyperbolic plane, $\Delta_g$.
The crucial difference from flat space is that the spectrum of $-\Delta_g$ begins at $|\kappa|/4$, rather than at zero~\cite{mckean1970,chavel1984,camporesi1994}. 
Consequently, the continuum antiferromagnetic magnon spectrum has an $O(S)$ bulk gap
\begin{equation}
    \Delta_{\rm AFM}
    =
    \frac{c}{2}\sqrt{|\kappa|}
    =
    \frac{JSq}{2\sqrt{2}}
    \sqrt{|\kappa|a^2}.
    \label{eq:AFM_continuum_gap_main}
\end{equation}
This is the continuum analogue of the lattice bulk gap extracted from the local density of states in Fig.~\ref{fig:gap}.

The same continuum theory also reproduces the exponential form of the equal-time transverse correlations. 
Specializing the calculation in App.~\ref{app:AFM_continuum} to dimension $d=2$ gives the long-distance asymptotic form
\begin{equation}
    |C_{\perp}(r)|
    \sim
    \frac{
        e^{-\sqrt{|\kappa|}\,r}
    }{
        \sqrt{r}
    }.
    \label{eq:AFM_continuum_correlation_main}
\end{equation}
This continuum result justifies the functional form of the lattice fit in Eq.~\eqref{eq:AFM_hyperbolic_transverse_fit}, with transverse correlation length equal to the radius of curvature
\begin{equation}
    \xi_{\perp}
    =
    \frac{1}{\sqrt{|\kappa|}}.
    \label{eq:AFM_continuum_correlation_length_main}
\end{equation}
The finite correlation length is therefore already present in the massless continuum Goldstone theory on negatively curved space~\cite{callan1990}. 
It is not generated by a spin anisotropy or by an explicit mass term, but rather by the intrinsic spectral gap of the Laplace--Beltrami operator.
For the $\{8,3\}$ lattice, Eq.~\eqref{eq:AFM_continuum_correlation_length_main} predicts a transverse correlation length $\xi_\perp/a \simeq 1.38$, compared with the fitted lattice value $\xi_\perp/a\simeq1.8$ in Fig.~\ref{fig:AFM_transverse_correlation}.
The fitted value is instead closer to the inverse graph-volume entropy $1/\ln\lambda_{8,3}\simeq 1.84$ of Eq.~\eqref{eq:8_3_volume_entropy}, suggesting that lattice growth rate provides the more accurate length scale even for this moderately curved lattice.

The ferromagnetic continuum limit is distinct from the antiferromagnetic one because the Berry-phase term is first order in imaginary time. 
After expanding about an ordered state and combining the two real transverse fields into a single complex field $\psi$, the quadratic continuum action takes the form
\begin{equation}
    S_{\rm FM}[\psi]
    =
    \frac{\rho_s}{JSqa^2}
    \int \dd\tau \,\dd z\,\dd\bar z\sqrt{|g|}
    \psi^*
    \left(
        \partial_\tau
        -
        D \Delta_g
    \right)
    \psi ,
    \label{eq:FM_continuum_action_main}
\end{equation}
with $D={|J|Sqa^2}/{4}$.
In this case, the magnon gap at $O(S)$ is
\begin{equation}
    \Delta_{\rm FM}
    =
    \frac{D|\kappa|}{4}
    =
    \frac{|J|Sq}{16}
    |\kappa|a^2 .
    \label{eq:FM_continuum_gap_main}
\end{equation}
The continuum curves in Fig.~\ref{fig:gap} show very good agreement with the lattice spin-wave gaps in the low-curvature regime. 
This agreement is expected, since when $|\kappa|a^2\ll1$, the lattice ``samples'' the hyperbolic plane on scales much shorter than the curvature radius, the adjacency Laplacian reduces to the Laplace--Beltrami operator~\cite{boettcher2020}, and the lower band edge is controlled by long-wavelength modes. 
In this regime, the only infrared scale is the curvature radius, so the continuum estimates in Eqs.~\eqref{eq:AFM_continuum_gap_main} and~\eqref{eq:FM_continuum_gap_main} capture the leading $O(S)$ dependence of the bulk gap.
For the ferromagnet, the ground state is a fully polarized product state, so transverse correlations trivially vanish for different sites at zero temperature.

When $|\kappa|a^2$ becomes large, the curvature radius is no longer well separated from the lattice spacing, and symmetry-allowed higher gradient terms are no longer suppressed relative to the leading Laplace--Beltrami operator~\cite{boettcher2020}.
The continuum theory should therefore be viewed as the controlled low-curvature limit of the lattice spin-wave theory, not as a description of the full high-curvature crossover.
Taken together, the lattice and continuum results give a consistent picture of curvature-induced infrared regularization, which we now summarize.

\section{Conclusion}
\label{sec:conclusion}

We have shown that hyperbolic geometry qualitatively reorganizes the infrared spectrum of the nearest-neighbor Heisenberg model. 
The ordered states retain the zero-energy Goldstone modes required by spontaneous breaking of $\mathrm{SO}(3)$ symmetry, but these modes are isolated from the bulk magnon continuum and carry vanishing local spectral weight in the thermodynamic bulk. 
Consequently, the relevant infrared scale for local observables is not the global symmetry mode itself, but the lower edge of the bulk magnon spectrum. 
For both the ferromagnet and the bipartite antiferromagnet, this edge is separated from zero frequency by a finite curvature-induced gap.

This provides a geometric route to circumvent the infrared instability that forbids finite-temperature magnetic long-range order in two-dimensional Euclidean Heisenberg magnets~\cite{mermin1966,hohenberg1967}. 
The mechanism does not require interactions that are long-range or that explicitly break $\mathrm{SO}(3)$ symmetry. 
Instead, it reflects the non-Euclidean structure of the underlying graph: Hyperbolic lattices support isolated global symmetry rotations without generating a locally visible continuum of arbitrarily soft magnons.
In spin-wave theory, the spectral gap renders the order-parameter correction finite even at nonzero temperatures.
In this sense, the usual Mermin--Wagner obstruction is absent for bulk hyperbolic lattices.

A second consequence is that negative curvature drives the spin-wave properties toward a more mean-field-like regime, manifested as a suppression of zero-point and thermal fluctuations. 
This interpretation is consistent with the dimensional crossover emphasized in recent work on hyperbolic systems: The geometry is locally two-dimensional on scales short compared with the curvature radius, while exponential volume growth makes the asymptotic long-distance problem effectively high- or infinite-dimensional~\cite{altland2026}. 
Related mean-field-like behavior has also been found in classical statistical models and continuum field theories on hyperbolic lattices and hyperbolic space~\cite{rietman1992,shima2006,krcmar2008,iharagi2010,gendiar2012,gendiar2014,breuckmann2020}. 
Our results show that this crossover is visible directly at the level of the magnon spectrum, as the curvature cuts off the low-energy phase space that would otherwise dominate in flat two dimensions.

The real-space transverse correlations reveal a complementary aspect of the same geometry. 
The connected equal-time transverse correlator is short ranged in graph distance, even though global integrated response functions can remain large or divergent because the number of sites grows exponentially with radius. 
Conversely, a divergent global response does not always imply the onset of long-range magnetic order, as in the case of the Bethe lattice Ising model, where the susceptibility diverges even in the absence of spontaneous average magnetization~\cite{matsuda1974,eggarter1974,wang2025}.
Thus local correlations and global response functions need not diagnose the same infrared physics. 
Motivated by the Bethe lattice result, the continuum calculation, and the numerical data for the $\{8,3\}$ lattice, we conjecture that the asymptotic transverse correlation length is set by the inverse graph-volume entropy. 
If generally valid, this relation gives a simple geometric interpretation of the apparent tension between short-ranged local correlations and a divergent staggered structure factor. 
A proof of this conjecture for general regular hyperbolic $\{p,q\}$ lattices remains an open problem. 
Related issues have appeared in studies of Ising models on hyperbolic lattices, where bulk correlations may remain finite or exponentially decaying even when boundary or shell-integrated observables display non-Euclidean scaling behavior~\cite{iharagi2010,okunishi2024}.

The continuum and Bethe lattice limits clarify the range of validity of this picture. 
At weak curvature, the lattice results are naturally connected to continuum Goldstone theories on the hyperbolic plane, where the Laplace--Beltrami spectrum has a curvature-controlled lower edge~\cite{mckean1970,chavel1984,camporesi1994}. 
At strong curvature, the relevant physics becomes increasingly graph-theoretic, and the Bethe lattice provides the limiting reference point at fixed coordination. 
Together, these limits show that the curvature-induced magnon gap is not a finite-size effect or a boundary artifact, but a bulk consequence of hyperbolic geometry.

Several directions follow directly from these results. 
First, the true finite-temperature transition should be determined beyond linear spin-wave theory, including magnon interactions, vortex or topological defects where appropriate, and critical fluctuations near the transition. 
Second, a rigorous theory of correlation functions on regular hyperbolic lattices would clarify when the inverse-volume-entropy conjecture holds and how local, shell-integrated, and boundary correlators are related. 
Third, finite hyperbolic flakes have extensive boundaries, so boundary-localized modes and boundary conditions may strongly affect spectra measured in finite samples even when the thermodynamic bulk is gapped. 
Separating bulk and boundary responses is therefore essential for comparing theory with synthetic hyperbolic platforms, including circuit-QED and topolectrical implementations~\cite{kollar2019,boettcher2020,bienias2022,zhang2022,lenggenhager2022,chen2023a,huang2024,chen2024a,yuan2025,xu2025}. 
Finally, the present work treats spins as internal $\mathrm{SO}(3)$ degrees of freedom with isotropic exchange. 
Including spin-orbit coupling, Dzyaloshinskii--Moriya interactions, anisotropic exchange, or coupling to local orbital frames would allow curvature to enter more directly through bond-dependent spin rotations and effective anisotropies. 
Such extensions would connect the internal-spin problem studied here to vector and tensor excitations, such as phonons and elastic modes, for which parallel transport is intrinsic to the long-wavelength dynamics.

\section*{Acknowledgements}
The authors acknowledge Igor Boettcher, Jongjun Lee, Frank Marsiglio, and Matthew Thomson for useful discussions. 
J.M. was supported by NSERC Discovery Grant No. RGPIN-2020-06999 and Quantum Horizons Alberta (QHA). 
This research was enabled in part by computing resources provided by the Digital Research Alliance of Canada (\href{https://www.alliancecan.ca/}{alliancecan.ca}).

\appendix
\numberwithin{equation}{section}
\numberwithin{figure}{section}

\section{Hyperbolic geometry and finite-lattice construction}
\label{appendix:HyperbolicLattice}

\begin{figure*}[t!]
    \centering
    \begin{overpic}[width=0.98\textwidth]{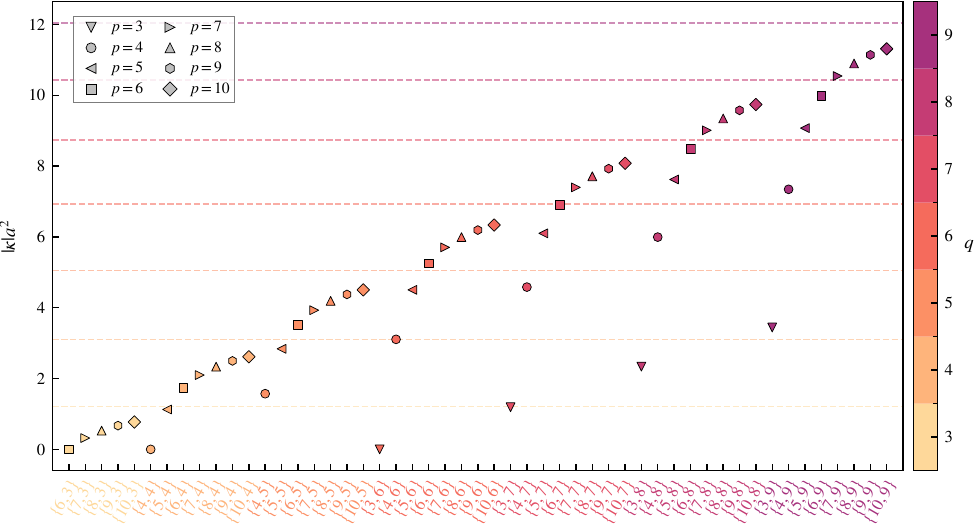}
    \end{overpic}
    \caption{
    Intrinsic curvature scale $|\kappa|a^2$ of Euclidean and hyperbolic $\{p,q\}$ lattices, computed from Eq.~\eqref{eq:pq_curvature_relation}.
    Symbols distinguish the number of sides $p$ of each polygonal face, while color denotes the coordination number $q$.
    The three Euclidean lattices $\{6,3\}$, $\{4,4\}$, and $\{3,6\}$ have zero curvature $|\kappa|a^2=0$.
    The dashed horizontal lines show the fixed-$q$ Bethe lattice values in Eq.~\eqref{eq:bethe_curvature_bound}, which give the $p\to\infty$ upper bounds within the $\{p,q\}$ family.
    }
    \label{fig:graph_curvature}
\end{figure*}

This appendix collects several geometric and numerical conventions used throughout the paper.
The main text defines the regular $\{p,q\}$ tilings, the Poincar\'e disk metric, the geodesic distance, and the curvature--edge-length relation in Eq.~\eqref{eq:pq_curvature_relation}.
In this appendix, we supplement that discussion by describing how finite open clusters are generated for the adjacency-matrix calculations, how the Poincar\'e disk coordinates are used to embed finite clusters, and how the intrinsic curvature scale varies across the hyperbolic $\{p,q\}$ lattices.
The shell-growth rate is discussed separately in App.~\ref{app:ShellGrowth}.

\subsection{Isometries} \label{appsubsec:DiskCoordinates}

We represent the hyperbolic plane by the Poincar\'e disk,
\begin{equation}
    \mathbb{D}=\{z\in\mathbb{C}: |z|<1\},
\end{equation}
with metric defined in Eq.~\eqref{eq:poincare_metric}.
The complex coordinate $z=x+iy$ is used to embed and visualize the lattice in hyperbolic space.
The spin Hamiltonian and the spin-wave spectra depend on the graph adjacency matrix, not on the particular disk embedding.
Thus two embeddings related by an isometry of the disk represent the same hyperbolic lattice.

The orientation-preserving isometries of the Poincar\'e disk can be written as the M\"obius transformations~\cite{magnus1974}
\begin{equation}
    z \mapsto 
    e^{i\theta}
    \frac{z-\zeta}{1-\bar{\zeta} z},
    \qquad \zeta \in \mathbb{D}.
    \label{eq:poincare_isometry}
\end{equation}
These transformations preserve the Poincar\'e disk metric in Eq.~\eqref{eq:poincare_metric}.
A coordinate-based construction of a hyperbolic $\{p,q\}$ lattice embedded in $\mathbb{D}$ begins with a regular $p$-gon centered at $z=0$, whose vertices are placed at equal angular intervals.
The radial position is then chosen so that adjacent vertices have geodesic distance $a$, fixed by Eq.~\eqref{eq:pq_curvature_relation}.
Neighboring polygons may be generated by applying disk isometries and reflections across existing geodesic edges.
This coordinate construction is useful for fixing the geometric interpretation of the lattice.
For the adjacency-matrix calculations discussed in Sec.~\ref{sec:model_methods}, however, we generate only the graph adjacency structure, using the vertex-inflation procedure described in Sec.~\ref{appsubsec:VertexInflation}.
Once the nearest-neighbor adjacency matrix has been generated, the vertex positions in the Poincar\'e disk no longer enter any observables.

\subsection{Curvature scale of a \texorpdfstring{$\{p,q\}$}{{p,q}} lattice} \label{appsubsec:CurvatureScale}

For a regular $\{p,q\}$ tiling, the edge length $a$ and Gaussian curvature $\kappa<0$ are related by Eq.~\eqref{eq:pq_curvature_relation}.
This expression makes explicit that, once the microscopic edge length $a$ is fixed, the curvature scale is determined entirely by the corresponding Schl\"afli symbol.
Figure~\ref{fig:graph_curvature} shows the resulting intrinsic curvature scale for some of the $\{p,q\}$ lattices considered in this work.
At fixed coordination number $q$, increasing $p$ leads to increasing $|\kappa|a^2$.
The formal $p\to\infty$ limit gives
\begin{equation}
    |\kappa|a^2
    =
    4\,\operatorname{arcosh}^2
    \left[
        \frac{1}{\sin(\pi/q)}
    \right],
    \qquad (p\to\infty)
    \label{eq:bethe_curvature_bound}
\end{equation}
which is shown by the dashed horizontal lines in Fig.~\ref{fig:graph_curvature}.
This limiting value provides an upper bound on the curvature scale at fixed $q$ within the $\{p,q\}$ family.
As discussed in Sec.~\ref{subsec:HyperbolicGeometry}, the $p\to\infty$ limit should be understood as the $q$-regular Bethe lattice.
In this limit, the elementary faces have become infinite-sided polygons, and the lattice has no finite closed cycles~\cite{mosseri1982,soderberg1993}.

The same curvature scale $|\kappa|a^2$ is used as the horizontal axis in Figs.~\ref{fig:gap}--\ref{fig:Tc} in the main text.
It provides a compact way to compare different hyperbolic lattices, but it should not be interpreted as an additional independent parameter of the lattice model.
Changing $|\kappa|a^2$ corresponds to changing the underlying $\{p,q\}$ lattice, and is therefore not a continuously tunable parameter.

\subsection{Vertex inflation and finite open clusters}
\label{appsubsec:VertexInflation}

The adjacency matrices used for numerical calculations in this work are generated from finite open clusters of the infinite $\{p,q\}$ tiling.
We use a vertex-inflation construction, closely related to the procedures used in Refs.~\cite{jahn2020,boyle2020,huang2025}.
The construction starts from a central regular $p$-gon.
The graph is then grown combinatorially by keeping track of the ordered outer boundary and repeatedly completing boundary vertices according to the local rule that $q$ regular $p$-gons meet at each vertex.

Each growth step begins with the current boundary of the finite cluster and identifies the boundary vertices whose local environment is not yet complete.
For every boundary edge incident to such a vertex, a new $p$-gon is attached on the exterior side of the cluster.
Since adjacent attachments can initially create duplicate boundary vertices, the construction then merges the vertices that represent the same site of the regular $\{p,q\}$ tiling.
After these local merges, boundary vertices whose incident $p$-gons have been completed are removed from the open set.
This attach--merge--complete procedure is repeated until all vertices from the current boundary layer have been completed.
The ordered outer boundary of the resulting cluster is then used as the starting boundary for the next growth step.
Rather than assigning coordinates to every new polygon in the Poincar\'e disk, this procedure constructs the same local $\{p,q\}$ adjacency structure directly from the boundary data.
The disk embedding introduced in Sec.~\ref{appsubsec:DiskCoordinates} is therefore used to define the geometric interpretation of the lattice and the curvature scale, whereas the numerical construction of the adjacency matrix $\vb*{A}$ is purely graph-theoretic.

Because the construction is terminated after a finite number of growth steps, there are boundary vertices that have fewer than $q$ neighbors.
Thus the finite adjacency matrix generated in this way is not a finite $q$-regular graph, but rather an open finite subgraph of the infinite $q$-regular bulk graph.
This distinction is important because finite hyperbolic clusters have a boundary whose size remains comparable to the total number of sites, so boundary effects need not disappear in the thermodynamic limit~\cite{breuckmann2020}.
Consequently, the full spectrum of an open finite cluster must not be interpreted as the finite-size spectrum of the thermodynamic bulk system.
In particular, boundary-localized eigenstates and coordination deficits significantly affect the spectrum of a finite cluster~\cite{chen2023a,tolosa-simeon2026}.
For this reason, finite clusters are used only as auxiliary objects for computing bulk \emph{local} quantities.
This use of large open hyperbolic clusters to infer the thermodynamic bulk density of states follows the strategy of Ref.~\cite{mosseri2023}.
The continued-fraction method used to extract this bulk local density of states is described in App.~\ref{appendix:LDOS}.

\subsection{Graph distance and geodesic distance}
\label{appsubsec:GraphDistance}

Two notions of distance appear in this work: the continuous geodesic distance from the Poincaré disk embedding used in Sec.~\ref{subsec:HyperbolicGeometry}, and the discrete graph distance counting nearest-neighbor bonds used in Sec.~\ref{subsec:transverse_correlations}.
The Poincar\'e disk embedding gives a continuous hyperbolic geodesic distance, Eq.~\eqref{eq:geodesic_distance}, between the embedded coordinates of two sites.
The graph itself also defines a discrete distance, equal to the minimum number of nearest-neighbor bonds connecting two vertices.
When distances are used in the spin correlation functions, we use the graph distance multiplied by the edge length $a$.
Thus, if two sites are separated by $n$ nearest-neighbor steps, we write their separation as $r=na$.

Using graph distance for lattice correlation functions is natural for the Heisenberg model in Eq.~\eqref{eq:HeisenbergModel}, because the Hamiltonian is defined by nearest-neighbor bonds of the graph.
The geodesic embedding fixes the intrinsic curvature scale and provides a geometric visualization of the lattice.
By contrast, the graph distance determines the shells of sites around a chosen bulk vertex and therefore sets the natural radial coordinate for graph-theoretic quantities.
The growth of these graph-distance shells is quantified in App.~\ref{app:ShellGrowth}.

\section{Shell-growth rate of a hyperbolic lattice}
\label{app:ShellGrowth}

In this appendix, we compute the exponential growth rate of graph-distance shells for a hyperbolic $\{p,q\}$ lattice.
This quantity is relevant to the real-space correlation function computed in Sec.~\ref{subsec:transverse_correlations}, where spatial separations are measured in graph distance rather than by the continuous geodesic distance of Eq.~\eqref{eq:geodesic_distance}.
The growth rate admits two equivalent asymptotic interpretations.
For the infinite lattice, it controls the number of vertices at graph distance $n$ from a fixed bulk vertex.
For the finite open clusters generated by the vertex-inflation construction of Sec.~\ref{appsubsec:VertexInflation}, it controls the large-$n$ growth of the number of boundary sites with the number of growth steps.
This equivalence is asymptotic as the outer boundary of a finite lattice depends on the specific details of its construction, whereas the exponential growth rate is an intrinsic property of the infinite $\{p,q\}$ graph.

Define the number of vertices at graph distance $n$ from a fixed bulk vertex as $\mathcal{Q}_n$, and define the shell-counting function of a $\{p,q\}$ lattice
\begin{equation}
    F_{p,q}(x)
    \equiv
    \sum_{n=0}^\infty \mathcal{Q}_n x^n .
    \label{eq:shell_counting_function}
\end{equation}
For a hyperbolic lattice, the shell-counting sequence grows exponentially at large $n$,
\begin{equation}
    \mathcal{Q}_n \sim \lambda^n ,
    \label{eq:shell_growth_asymptotic}
\end{equation}
for some growth rate $\lambda>1$.
Equivalently, the number of boundary sites in a large finite open cluster grows with the same exponential rate.
The Cauchy--Hadamard formula~\cite{lang1999} relates the growth rate in Eq.~\eqref{eq:shell_growth_asymptotic} to the radius of convergence of Eq.~\eqref{eq:shell_counting_function},
\begin{equation}
    \lambda = \limsup_{n\to \infty} \mathcal{Q}_n^{1/n}
    =
    \frac{1}{\mathcal{R}},
    \label{eq:cauchy_hadamard_shell}
\end{equation}
where $\mathcal{R}$ is the radius of convergence of $F_{p,q}(x)$.
The corresponding \emph{volume entropy} is
\begin{equation}
    \ln \lambda = -\ln \mathcal{R} .
    \label{eq:volume_entropy}
\end{equation}
In units of the edge length, the inverse volume entropy defines the geometric graph-distance scale $a/\ln \lambda$.

Refs.~\cite{floyd1994,bartholdi2002} list the shell-counting function $F_{p,q}(x)$ for a regular $\{p,q\}$ lattice as a rational function.
The explicit form depends on the parity of $p$.
For $p=2w$,
\begin{equation}
    F_{2w,q}(x) =
    \frac
    {\displaystyle 1+2\sum_{r=1}^{w-1}x^r+x^w}
    {\displaystyle 1-(q-2)\sum_{r=1}^{w-1}x^r+x^w}.
    \label{eq:shell_generating_even}
\end{equation}
For $p=2w+1$,
\begin{equation}
    F_{2w+1,q}(x)=
    \frac{
    \displaystyle 1+2\sum_{r=1}^{2w-1}x^r+2x^w+x^{2w}
    }{
    \displaystyle 1-(q-2)\sum_{r=1}^{2w-1}x^r+2x^w+x^{2w}
    }.
    \label{eq:shell_generating_odd}
\end{equation}
Because the coefficients $\mathcal{Q}_n$ are nonnegative, the Vivanti--Pringsheim theorem states that the point $x=\mathcal{R}$ is a singularity of $F_{p,q}(x)$~\cite{remmert1991}.
For the rational functions in Eqs.~\eqref{eq:shell_generating_even} and~\eqref{eq:shell_generating_odd}, this singularity occurs at the smallest positive root of the denominator.
Equivalently, the denominator defines the characteristic polynomial of a finite recurrence relation for the shell-counting problem.
Its dominant root controls the large-distance exponential growth of graph-distance shells.
In practice, we compute the growth rate by finding the smallest positive root $\mathcal{R}$ of the denominator and using Eq.~\eqref{eq:cauchy_hadamard_shell}.

For $p=3-6,8,$ and $10$, we find exact solutions:
\begin{subequations}
\label{eq:lambdas}
\begin{align}
    \lambda_{3,q} &=
    \frac{1}{2} \left[(q-4) + \sqrt{(q-4)^2-4} \right], \\
    \lambda_{4,q} &=
    \frac{1}{2} \left[(q-2) + \sqrt{(q-2)^2-4} \right] ,\\
    \lambda_{5,q} &=
    \frac{1}{4} \Biggl[(q-2) + \sqrt{q^2-4} 
    \nonumber \\
    &\qquad\qquad\qquad+ \sqrt{\left( q-2 + \sqrt{q^2-4} \right)^2 - 16} \Biggr], \\
    \lambda_{6,q} &=
    \frac{1}{2} \left[(q-1) + \sqrt{(q-1)^2-4} \right], \\
    \lambda_{8,q} &=
    \frac{1}{4} \Biggl[(q-2) + \sqrt{q^2+4} \nonumber \\
    &\qquad\qquad\qquad+ \sqrt{\left( q-2 + \sqrt{q^2+4} \right)^2 - 16} \Biggr], \\
    \lambda_{10,q} &=
    \frac{1}{4} \Biggl[(q-1) + \sqrt{(q-1)^2+4} \nonumber \\
    &\qquad\qquad+ \sqrt{\left( q-1 + \sqrt{(q-1)^2+4} \right)^2 - 16} \Biggr],
\end{align}
\end{subequations}
where $\lambda_{p,q}$ denotes the growth rate of the corresponding $\{p,q\}$ lattice.
For the remaining values of $p,q$ not listed in Eqs.~\eqref{eq:lambdas}, the growth rate can be obtained numerically by finding the smallest root of the denominator of either Eq.~\eqref{eq:shell_generating_even} or Eq.~\eqref{eq:shell_generating_odd}.
In the Bethe lattice limit $p\to \infty$, the denominator of Eqs.~\eqref{eq:shell_generating_even} and~\eqref{eq:shell_generating_odd} can be evaluated as a power series, corresponding to a growth rate $\lambda_{\infty,q}=q-1$.
This gives the graph-distance scale $a/\ln(q-1)$ that appears in the exact Bethe lattice transverse-correlation asymptotics discussed in App.~\ref{app:BetheLattice}.

\begin{figure}[tbp]
    \centering
    \begin{overpic}[width=\columnwidth]{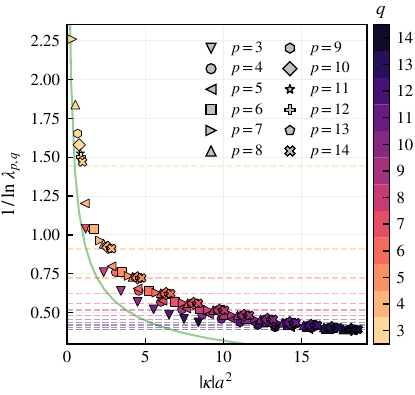}
    \end{overpic}
    \caption{
    Inverse volume entropy $1/\ln\lambda_{p,q}$ of regular hyperbolic $\{p,q\}$ lattices as a function of the intrinsic curvature scale $|\kappa|a^2$.
    Symbols distinguish $p$, while color denotes $q$.
    The dashed horizontal lines indicate the Bethe lattice values $1/\ln(q-1)$ at fixed $q$.
    The solid green curve shows the continuum geometric scale $1/\sqrt{|\kappa|a^2}$.
    }
    \label{fig:inverse_volume_entropy}
\end{figure}

Figure~\ref{fig:inverse_volume_entropy} shows the inverse volume entropy as a function of the intrinsic curvature scale $|\kappa|a^2$ defined in Eq.~\eqref{eq:pq_curvature_relation}.
At fixed $q$, the inverse volume entropy is bounded from below by the Bethe lattice limit $1/\ln(q-1)$, depicted by the dashed horizontal lines in Fig.~\ref{fig:inverse_volume_entropy}.
The inverse volume entropy tends to decrease as the curvature scale increases, reflecting the faster exponential growth of graph-distance shells in more strongly curved lattices.
The solid green curve shows the continuum geometric scale $1/\sqrt{|\kappa|a^2}$ for comparison.

The quantity plotted in Fig.~\ref{fig:inverse_volume_entropy} is \emph{not} the transverse spin correlation length extracted from Eq.~\eqref{eq:AFM_hyperbolic_transverse_fit}.
For the Bethe lattice, the exact asymptotic result does indeed give a transverse correlation length equal to the inverse volume entropy in units of $a$, namely $\xi_\perp/a=1/\ln(q-1)$.
However, for finite-$p$ hyperbolic lattices, Fig.~\ref{fig:inverse_volume_entropy} shows only a geometric quantity of the lattice, the inverse volume entropy $1/\ln \lambda_{p,q}$.
Identifying this quantity with the transverse correlation length $\xi_\perp/a$ would require an additional argument beyond the shell-growth analysis presented in this appendix, and remains an open question.

\section{Local density of states calculation}
\label{appendix:LDOS}

In this appendix, we describe the continued-fraction method used to compute the bulk local density of states of the adjacency matrix.
This is the central numerical input for the local magnon density of states and the bulk thermodynamic quantities presented in Sec.~\ref{sec:results}.
The finite clusters used in the calculation are the open hyperbolic clusters generated by the vertex-inflation procedure described in Sec.~\ref{appsubsec:VertexInflation}.
As discussed in Sec.~\ref{appsubsec:VertexInflation}, the full spectrum of such an open cluster contains boundary contributions and should not be interpreted as the thermodynamic bulk spectrum.
We therefore compute the local spectral measure from a site chosen deep in the interior of the cluster.
This construction follows the real-space continued-fraction method of Refs.~\cite{haydock1972,haydock1975,viswanath1994,mosseri2023,lenggenhager2025}.

\subsection{Adjacency local Green's function}

Let $\vb*{A}$ denote the adjacency matrix of the $\{p,q\}$ lattice.
For a site-localized basis state $\ket{i}$, we define the local Green's function of the adjacency matrix as
\begin{equation}
    g_{ii}(z)
    =
    \mel{i}{\left(z-\vb*{A}\right)^{-1}}{i}.
    \label{eq:adjacency_local_green_function}
\end{equation}
The corresponding adjacency local density of states is
\begin{equation}
    \nu_i(\lambda)
    =
    -\frac{1}{\pi}
    \Im g_{ii}(\lambda+i0^+).
    \label{eq:adjacency_ldos_from_green}
\end{equation}
This is the local spectral measure that enters the spin-wave expressions in Sec.~\ref{sec:model_methods}.
For the infinite regular $\{p,q\}$ lattice, all bulk sites are equivalent, so the thermodynamic local density of states is independent of the reference site.
For a large finite open cluster, this bulk quantity is approximated by choosing $i$ near the center of the cluster.

\subsection{Continued-fraction method}
\label{subsec:continued-fraction}

The continued-fraction method is a Lanczos recursion initialized with the local state $\ket{i}$.
We set
\begin{equation}
    \ket{\varphi_0}=\ket{i},
    \qquad
    \ket{\varphi_{-1}}=0,
    \qquad
    \beta_0=0 .
\end{equation}
For $n=0,1,2,\ldots$, define
\begin{align}
    \alpha_n
    &=
    \mel{\varphi_n}{\vb*{A}}{\varphi_n},
    \label{eq:lanczos_alpha}
    \\
    \ket{\tilde{\varphi}_{n+1}}
    &=
    \vb*{A}\ket{\varphi_n}
    -
    \alpha_n\ket{\varphi_n}
    -
    \beta_n\ket{\varphi_{n-1}},
    \label{eq:lanczos_vector}
    \\
    \beta_{n+1}
    &=
    \sqrt{
    \braket{\tilde{\varphi}_{n+1}}{\tilde{\varphi}_{n+1}}
    },
    \label{eq:lanczos_beta}
    \\
    \ket{\varphi_{n+1}}
    &=
    \frac{1}{\beta_{n+1}}
    \ket{\tilde{\varphi}_{n+1}} .
    \label{eq:lanczos_normalization}
\end{align}
The recursion terminates only if $\beta_{n+1}=0$.
Otherwise, it generates an orthonormal Krylov basis in which the adjacency matrix is tridiagonal,
\begin{equation}
    \vb*{A}
    \;\longrightarrow\;
    \begin{pmatrix}
        \alpha_0 & \beta_1 & 0 & 0 & 0 & \cdots \\
        \beta_1 & \alpha_1 & \beta_2 & 0 & 0 & \cdots \\
        0 & \beta_2 & \alpha_2 & \beta_3 & 0 & \cdots \\
        0 & 0 & \beta_3 & \alpha_3 & \beta_4 & \cdots \\
         0 & 0 & 0 & \beta_4 & \alpha_4 & \cdots \\
        \vdots & \vdots & \vdots & \vdots & \vdots &\ddots \\
    \end{pmatrix}.
    \label{eq:lanczos_tridiagonal_matrix}
\end{equation}
Because the initial vector is $\ket{i}$, the local Green's function in Eq.~\eqref{eq:adjacency_local_green_function} is the upper-left matrix element of the resolvent of Eq.~\eqref{eq:lanczos_tridiagonal_matrix}.
This gives the continued fraction representation
\begin{equation}
    g_{ii}(z)
    =
    \frac{1}{
    z-\alpha_0
    -
    \dfrac{\beta_1^2}{
    z-\alpha_1
    -
    \dfrac{\beta_2^2}{
    z-\alpha_2-\cdots
    }}
    } .
    \label{eq:adjacency_continued_fraction}
\end{equation}
For a bipartite lattice, the adjacency spectrum is symmetric, $\nu_i(\lambda) = \nu_i(-\lambda)$.
In this case, the recursion initialized on a single site has $\alpha_n=0$ for all $n$, and the continued fraction contains only the off-diagonal coefficients $\beta_n$.

Each application of $\vb*{A}$ increases the graph-distance support of a Krylov vector by at most one.
Consequently, the first $n$ recursion coefficients depend only on the graph structure within distance $n$ of the reference site.
For a reference site $i$ in the bulk of a large finite lattice, these coefficients are equivalent to the infinite-lattice coefficients until the recursion reaches the boundary.
Increasing the number of vertex-inflation growth steps therefore increases the number of boundary-independent recursion coefficients.
This is the sense in which open hyperbolic clusters are used as auxiliary objects for computing a thermodynamic bulk local spectral measure.

In numerical calculations, the continued fraction is evaluated after a finite number of recursion steps.
Equivalently, we write
\begin{equation}
    g_{ii}^{(M)}(z)
    =
    \frac{1}{
    z-\alpha_0
    -
    \dfrac{\beta_1^2}{
    z-\alpha_1
    -
    \dfrac{\beta_2^2}{
    \ddots
    -
    \beta_M^2 t_M^{}(z)
    }}
    },
    \label{eq:finite_continued_fraction}
\end{equation}
where $t_M(z)$ denotes the tail of the continued fraction.
A direct finite-depth approximation corresponds to choosing the tail by truncation and evaluating the result at $z=\lambda+i\eta$ with small $\eta \sim 10^{-4}$.
If the recursion coefficients reach a stable asymptotic regime, the tail may instead be replaced by an analytic terminator, as in the standard recursion method~\cite{haydock1972,haydock1975,viswanath1994}.
For example, if
\begin{equation}
    \alpha_n\to \alpha_\infty,
    \qquad
    \beta_n\to \beta_\infty,
    \label{eq:lanczos_asymptotic_coefficients}
\end{equation}
then the asymptotic tail satisfies
\begin{equation}
    t(z)
    =
    \frac{1}{
    z-\alpha_\infty-\beta_\infty^2 t(z)
    } .
    \label{eq:constant_terminator_definition}
\end{equation}
Solving Eq.~\eqref{eq:constant_terminator_definition} gives
\begin{equation}
    t(z)
    =
    \frac{
    z-\alpha_\infty
    -
    \sqrt{(z-\alpha_\infty)^2-4\beta_\infty^2}
    }{
    2\beta_\infty^2
    } ,
    \label{eq:constant_terminator_solution}
\end{equation}
where the physical branch of the square root is chosen so that $t(z)\sim 1/z$ as $|z|\to\infty$.
The local density of states is then obtained from Eq.~\eqref{eq:adjacency_ldos_from_green}.

\subsection{Ferromagnetic magnon density of states}

For the Heisenberg ferromagnet, the quadratic spin-wave Hamiltonian is diagonal in the adjacency eigenbasis.
The magnon frequency associated with an adjacency eigenvalue $\lambda$ is given by Eq.~\eqref{eq:FM_solution}.
Changing variables from $\lambda$ to $\omega$ gives the local ferromagnetic magnon density of states
\begin{equation}
    \rho_i^{\rm FM}(\omega)
    =
    \frac{1}{S|J|}
    \nu_i\!\left(
        q-\frac{\omega}{S|J|}
    \right).
    \label{eq:FM_dos_from_adjacency_appendix}
\end{equation}
The isolated uniform mode Eq.~\eqref{eq:uniform_eigenvector} at $\lambda=q$ maps to $\omega=0$.
On a finite $q$-regular graph this mode has local weight $1/N$, therefore it does not contribute to the thermodynamic bulk local density of states at a fixed site.

\begin{figure*}[p]
    \centering
    \begin{overpic}[width=0.98\textwidth]{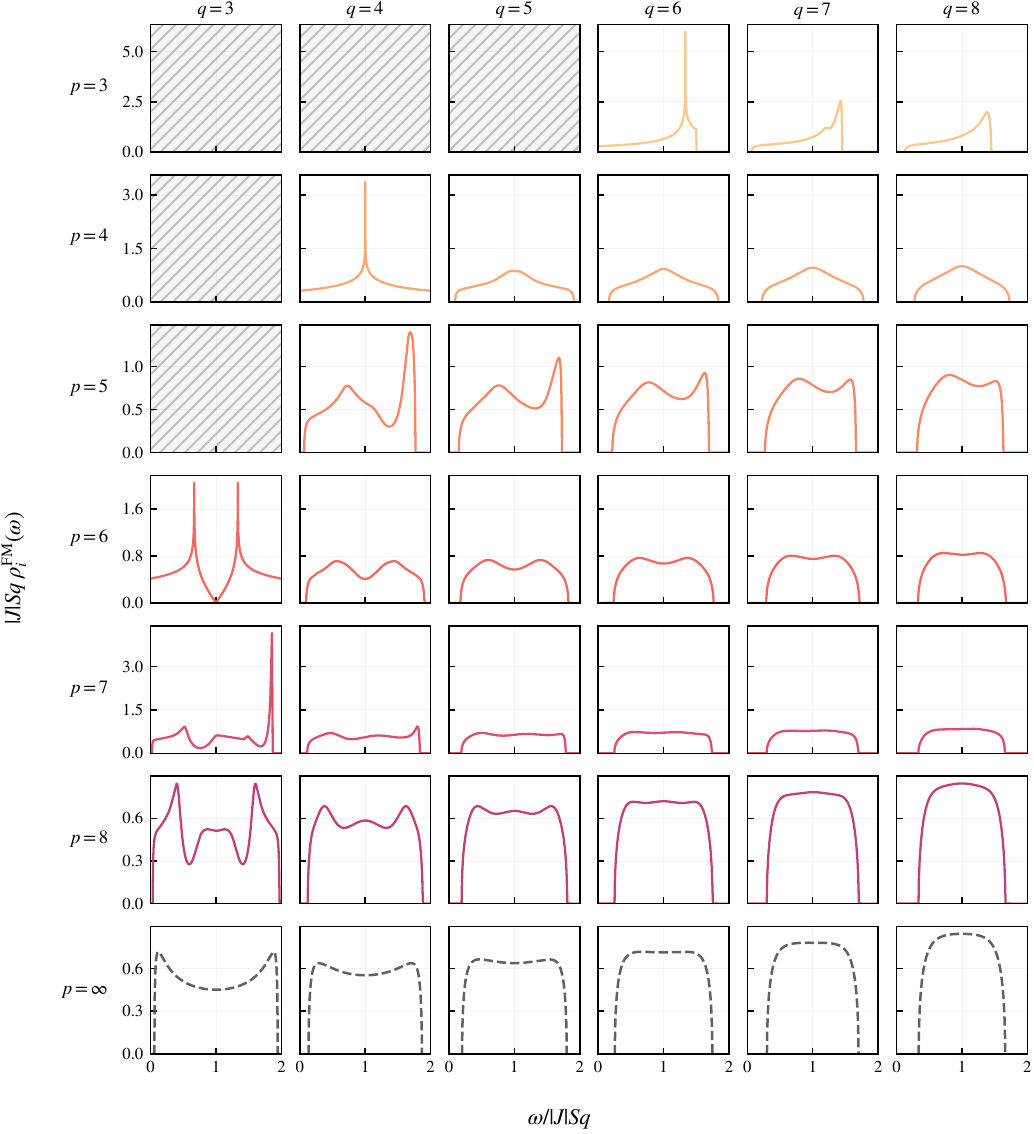}
    \end{overpic}
    \caption{
    Local density of states at $O(S)$ for the Heisenberg ferromagnet, obtained from the adjacency local density of states through Eq.~\eqref{eq:FM_dos_from_adjacency_appendix}.
    Hatched panels correspond to $\{p,q\}$ values outside the hyperbolic or Euclidean regimes.
    The $p=\infty$ row shows the Bethe lattice result.
    }
    \label{fig:FM_DOS}
\end{figure*}
Figure~\ref{fig:FM_DOS} shows the resulting ferromagnetic local density of states for some of the $\{p,q\}$ lattices considered in this work.
The plotted spectra are obtained by shifting and rescaling the adjacency local density of states according to Eq.~\eqref{eq:FM_dos_from_adjacency_appendix}.
The panels illustrate how negative curvature shifts spectral weight away from the low-energy continuum, opening the bulk magnon gap.
This gap is the same as the curvature-induced spectral gap depicted in Fig.~\ref{fig:gap}.

\subsection{Antiferromagnetic magnon density of states}

For the bipartite Heisenberg antiferromagnet, the quadratic spin-wave Hamiltonian is diagonalized by a bosonic Bogoliubov transformation.
The normal-mode frequencies are again determined by the adjacency eigenvalues.
For an adjacency eigenvalue $\lambda$, the antiferromagnetic spin-wave frequency is given by Eq.~\eqref{eq:AFM_solution}.
For a positive frequency $0<\omega<JSq$, define
\begin{equation}
    \lambda_\omega
    =
    \sqrt{
        q^2-
        \left(
            \frac{\omega}{JS}
        \right)^2
    } .
    \label{eq:lambda_omega_definition}
\end{equation}
The two adjacency eigenvalues $\lambda=\pm\lambda_\omega$ contribute to the same antiferromagnetic magnon frequency.
The local antiferromagnetic magnon density of states is therefore
\begin{equation}
    \rho_i^{\rm AFM}(\omega)
    =
    \frac{\omega}{(JS)^2\lambda_\omega}
    \left[
        \nu_i(\lambda_\omega)
        +
        \nu_i(-\lambda_\omega)
    \right].
    \label{eq:AFM_dos_from_adjacency_general_appendix}
\end{equation}
For the bipartite lattices considered in the antiferromagnetic calculation, $\nu_i(\lambda)=\nu_i(-\lambda)$.
Equation~\eqref{eq:AFM_dos_from_adjacency_general_appendix} then reduces to Eq.~\eqref{eq:AFM_dos_from_adjacency}.
As in the ferromagnetic case, the isolated Goldstone modes at $\lambda=\pm q$ carry only $O(1/N)$ local spectral weight on a finite graph.
They are therefore excluded from the thermodynamic bulk local density of states at a fixed site.
Equation~\eqref{eq:AFM_dos_from_adjacency} is the transformation used to generate the magnon local density and the thermodynamic quantities computed in Sec.~\ref{sec:results}.

\begin{figure}[t!]
    \centering
    \begin{overpic}[width=\columnwidth]{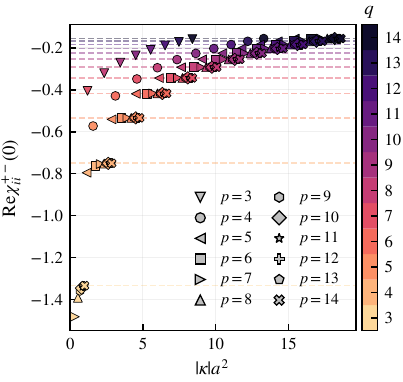}
    \end{overpic}
    \caption{
    Static local transverse susceptibility $\Re\chi_{ii}^{+-}(0)$ as a function of the intrinsic curvature scale $|\kappa|a^2$.
    The response remains finite for every hyperbolic lattice shown.
    For the bipartite lattices, the ferromagnetic and antiferromagnetic results are identical because the adjacency local density of states is symmetric under $\lambda\to-\lambda$.
    }
    \label{fig:ReChi0}
\end{figure}

Figure~\ref{fig:ReChi0} shows the static local transverse susceptibility computed from this bulk local spectral measure, given by Eq.~\eqref{eq:real_chi_from_dos}.
The susceptibility remains finite for all hyperbolic lattices shown, consistent with the finite low-frequency gap in the local magnon spectrum.
Its magnitude is reduced as the intrinsic curvature scale $|\kappa|a^2$ increases, reflecting the suppression of low-frequency local spectral weight.
For bipartite lattices, the ferromagnetic and antiferromagnetic values coincide because the adjacency local density of states is symmetric under $\lambda\to-\lambda$.
This agreement is therefore a consequence of the spectral symmetry of the underlying graph, not an additional assumption about the spin-wave calculation.

\subsection{Band-edge scaling and moment asymptotics}
\label{subsec:Band-edge_scaling}

The low-temperature thermodynamics discussed in Sec.~\ref{subsec:finite_temperature_transition} is controlled by the lower band edge of the bulk magnon continuum and, in particular, by the power-law exponent with which the local density of states vanishes at that edge.
For the continued fraction method described in App.~\ref{subsec:continued-fraction}, the constant tail terminator in Eq.~\eqref{eq:constant_terminator_solution} contains a square-root branch point and therefore produces a square-root lower band edge
\begin{equation}
    \rho_i(\omega) \sim |\omega - \Delta|^{1/2},\qquad \omega \to \Delta^+,
\end{equation}
where $\omega = \Delta$ is the lower band edge of the magnon continuum.
This raises an important consistency question. Since the terminator function in Eq.~\eqref{eq:constant_terminator_solution} is a numerical approximation to the asymptotics of the exact continued fraction, one should verify that the square-root edge exponent is not an artifact of this approximation.
A direct way to compute the edge exponent is to analyze the asymptotic form of the adjacency density of states moment, defined as
\begin{equation}
\mu_n
\equiv
\expval{\vb*{A}^n}{i}
=
\int \dd\lambda \,
\lambda^n \, \nu_i(\lambda).
\label{eq:DOS_moment}
\end{equation}
Assume that the local density of states has compact support $\lambda\in[\lambda_-,\lambda_+]$, and that near the two band edges it has the leading power-law form
\begin{equation}
\nu_i(\lambda)
\sim
A_\pm \left| \lambda-\lambda_\pm\right|^{\alpha_\pm},
\qquad
\lambda - \lambda_\pm \to 0^\mp ,
\end{equation}
with $A_\pm>0$ and $\alpha_\pm>-1$.
The condition $\alpha_\pm>-1$ ensures that the local density of states remains integrable. 
Without loss of generality, assume that the upper band edge is equal to or larger in magnitude $|\lambda_+| \geq |\lambda_-|$.
Next, choose a small parameter $\delta > 0$, such that 
\begin{equation}
    |\lambda| \leq |\lambda_+| - \eta, \quad \textrm{for all }\lambda \in [\lambda_-+\delta,\lambda_+-\delta],
\end{equation}
for some $\eta > 0$.
We split the moment into the sum of three terms $\mu_n^{} = \mu_n^{\rm{edge},-}+\mu_n^{\rm{edge},+}+\mu_n^{\rm{bulk}}$ with
\begin{align}
    \mu_n^{\rm{edge},-} &\equiv
    \int_{\lambda_-}^{\lambda_- + \delta} \dd{\lambda} \, \lambda^n \,\nu_i(\lambda),
    \\
    \mu_n^{\rm{edge},+} &\equiv
    \int_{\lambda_+-\delta}^{\lambda_+} \dd{\lambda} \, \lambda^n \,\nu_i(\lambda),
    \\
    \mu_n^{\rm{bulk}} &\equiv
    \int_{\lambda_-+\delta}^{\lambda_+-\delta} \dd{\lambda} \, \lambda^n \,\nu_i(\lambda).
\end{align}
To simplify notation, we introduce the band edge index $\sigma = \pm$ and the shifted eigenvalue $x \equiv \sigma (\lambda_\sigma - \lambda)$ to write the two edge contributions as
\begin{equation}
    \mu_n^{\rm{edge},\sigma} =
    \lambda_{\sigma}^n
    \int_{0}^{\delta} \dd{x} \, \left(1-\frac{x}{|\lambda_{\sigma}|}\right)^n \,\nu_i(\lambda_{\sigma}-\sigma x),
    \label{eq:moment_compact}
\end{equation}
and evaluate them in the asymptotic limit $n\to\infty$.
The precise statement of power-law scaling at the band edge is that, for any $\epsilon >0$, one can choose $\delta>0$ such that 
\begin{equation}
    A_\sigma (1-\epsilon) x^{\alpha_\sigma}
    \leq
    \nu_i(\lambda_{\sigma}-\sigma x)
    \leq
    A_\sigma (1+\epsilon) x^{\alpha_\sigma},
\end{equation}
for $0<x<\delta$.
It follows from Eq.~\eqref{eq:moment_compact} that
\begin{align}
    A_\sigma (1-\epsilon) \int_0^\delta &\dd{x} \, x^{\alpha_\sigma}
    \left(1-\frac{x}{|\lambda_{\sigma}|}\right)^n
    \leq
    \frac{\mu_n^{\rm{edge},\sigma}}{\lambda_\sigma^n} \nonumber \\
    &\leq
    A_\sigma (1+\epsilon) \int_0^\delta \dd{x} \, x^{\alpha_\sigma}
    \left(1-\frac{x}{|\lambda_{\sigma}|}\right)^n.
    \label{ineq:edge_moment}
\end{align}
Taking the limit $n\to \infty$ and making the substitution $y = nx/|\lambda_\sigma|$, we find
\begin{align}
    A_\sigma (1\pm\epsilon) &\int_0^\delta \dd{x} \, x^{\alpha_\sigma}
    \left(1-\frac{x}{|\lambda_{\sigma}|}\right)^n \nonumber \\
    &\to
    A_\sigma (1\pm\epsilon) |\lambda_{\sigma}|^{\alpha_\sigma+1}
    n^{-(\alpha_\sigma+1)}
    \int_0^\infty \dd{y} \ y^{\alpha_\sigma} e^{-y} \nonumber \\
    &= A_\sigma (1\pm\epsilon) |\lambda_{\sigma}|^{\alpha_\sigma+1}
    n^{-(\alpha_\sigma+1)} \Gamma\left(\alpha_\sigma + 1\right).
\end{align}
The inequality~\eqref{ineq:edge_moment} along with the squeeze theorem $(\epsilon\to0)$ thus implies the following asymptotic form of $\mu_n^{\rm{edge},\sigma}$
\begin{equation}
    \mu_n^{\rm{edge},\sigma} \sim A_\sigma |\lambda_{\sigma}|^{\alpha_\sigma+1}
    \Gamma\left(\alpha_\sigma + 1\right)
    \lambda_\sigma^n \,
    n^{-(\alpha_\sigma+1)}.
\end{equation}
Next, we find the following bound for the remaining contribution to the density of states moment $\mu_n^{\rm{bulk}}$
\begin{align}
    |\mu_n^{\rm{bulk}}|
    \leq
    \left(|\lambda_+|-\eta\right)^n
    \int \dd{\lambda} \, \nu_i(\lambda)
    &\leq
    \left(|\lambda_+|-\eta\right)^n \nonumber \\
    &=
    |\lambda_+|^n\left(1-\frac{\eta}{|\lambda_+|}\right)^n.
\end{align}
It follows that
\begin{align}
    |\mu_n^{\rm{bulk}}|
    \leq
    \frac{n^{\alpha_++1} \left(1-\frac{\eta}{|\lambda_+|}\right)^n}
    {A_+ |\lambda_{+}|^{\alpha_++1}
    \Gamma\left(\alpha_+ + 1\right)}
    |\mu_n^{\rm{edge},+}|
    \to 0,
\end{align}
as $n\to\infty$. 
That is, the bulk contribution is exponentially suppressed for large $n$.
The asymptotic form of $\mu_n$ therefore comes solely from the band edge contributions.

When $p$ is even, the graph is bipartite, and the density of states is symmetric $\nu_i(\lambda) = \nu_i(-\lambda)$~\cite{mosseri2023} (implying $\lambda_- = -\lambda_+$, $\alpha_- = \alpha_+$, etc.), so both band edges contribute equally to the asymptotic moment
\begin{equation}
    \mu_n \sim A_+ |\lambda_{+}|^{\alpha_++1}
    \Gamma\left(\alpha_+ + 1\right)
    \left(1+(-1)^n\right)
    \lambda_+^n \,
    n^{-(\alpha_++1)}.
    \label{eq:moment_edge_bipartite_asymptotic}
\end{equation}
When $p$ is odd, the graph is not bipartite, and the inequality $|\lambda_+| > |\lambda_-|$ is generally strict. In this case,
\begin{equation}
    \frac
    {|\mu_n^{\rm{edge},-}|}
    {|\mu_n^{\rm{edge},+}|}
    =
    \frac
    {A_- |\lambda_{-}|^{\alpha_-+1}\Gamma\left(\alpha_- + 1\right)}
    {A_+ |\lambda_{+}|^{\alpha_++1}\Gamma\left(\alpha_+ + 1\right)}
    \left(\frac{|\lambda_-|}{|\lambda_+|}\right)^n
    n^{\alpha_+ - \alpha_-}
    \to 0,
\end{equation}
so the lower band-edge contribution is suppressed. Therefore the asymptotic moment comes solely from the upper (largest-magnitude) band-edge
\begin{equation}
    \mu_n \sim A_+ |\lambda_{+}|^{\alpha_++1}
    \Gamma\left(\alpha_+ + 1\right)
    \lambda_+^n \,
    n^{-(\alpha_++1)}.
    \label{eq:moment_edge_asymptotic}
\end{equation}
Equations~\eqref{eq:moment_edge_bipartite_asymptotic} and \eqref{eq:moment_edge_asymptotic} show that a band-edge exponent $\alpha_+$ produces an asymptotic moment proportional to $n^{-(\alpha_++1)}$.
For the hyperbolic $\{p,q\}$ lattices, it was previously shown that $\mu_n\sim n^{-3/2}$~\cite{mckay1981,gouezel2013}, thus the bulk local density of states has a square-root band edge $\alpha_+ = 1/2$.
This provides an independent consistency check that the square-root edge obtained from the constant tail terminator in Eq.~\eqref{eq:constant_terminator_solution} is not merely an artifact of the approximation.

Note that the transformations between the adjacency eigenvalues and the magnon frequencies in Eqs.~\eqref{eq:FM_dos_from_adjacency_appendix} and~\eqref{eq:AFM_dos_from_adjacency_general_appendix} are smooth, and map the upper adjacency band edge to the lower magnon band edge.
Therefore the square-root adjacency band edge is inherited by the local magnon density of states,
\begin{equation}
\rho_i(\omega)
\sim
|\omega-\Delta|^{1/2},
\qquad
\omega\to\Delta^+ .
\end{equation}

\section{Exact solution for the Bethe lattice}
\label{app:BetheLattice}

\begin{figure}[t!]
\centering
\begin{overpic}[width=\columnwidth]{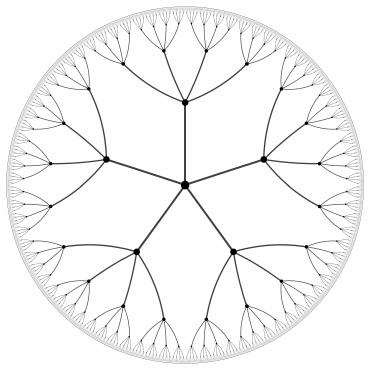}
\end{overpic}
\caption{
The $q=5$ Bethe lattice in the Poincar\'e-disk representation.
The Bethe lattice is the $\{p\to \infty,q\}$ limit of the regular $\{p,q\}$ hyperbolic lattices. 
}
\label{fig:BetheLattice}
\end{figure}

The Bethe lattice is the $p\to\infty$ limit of the $\{p,q\}$ lattices discussed in Sec.~\ref{subsec:HyperbolicGeometry}.
A finite-depth portion of the $q=5$ Bethe lattice is shown in Fig.~\ref{fig:BetheLattice} using a Poincar\'e-disk representation, emphasizing its interpretation as the $\{\infty,q\}$ hyperbolic tiling limit~\cite{mosseri1982,soderberg1993}.
It is an infinite connected graph in which every vertex has coordination number $q$ and there is a unique shortest path between any two vertices.
Because the graph has no finite closed loops, the adjacency Green's function and the corresponding spin-wave observables can be obtained analytically.
These results provide the $p=\infty$ reference values shown in Figs.~\ref{fig:AFM_LDOS}, \ref{fig:gap}--\ref{fig:Tc}, \ref{fig:FM_DOS}, and~\ref{fig:ReChi0}.

\subsection{Bethe lattice Green's function}
\label{app:Bethe_green_function}

We first derive the adjacency resolvent of the Bethe lattice.
Let
\begin{equation}
g_{ij}(z)
\equiv
\bra{i}
(z-\vb*{A})^{-1}
\ket{j},
\label{eq:Bethe_resolvent_definition}
\end{equation}
where $\vb*{A}$ is the adjacency matrix.
For fixed reference site $j$, the matrix element $g_{ij}(z)$ depends only on the graph distance $n$, which we denote by $g_n(z)$.
The equation
\begin{equation}
(z-\vb*{A})\vb*{g}(z)=\mathbbm{1}
\end{equation}
then gives
\begin{align}
z g_0(z)-q g_1(z) &= 1, \label{eq:Bethe_resolvent_recursion_0} \\
z g_n(z)-g_{n-1}(z)-(q-1)g_{n+1}(z) &=0,
\qquad n\geq 1.
\label{eq:Bethe_resolvent_recursion_n}
\end{align}
The second equation admits the solution~\cite{bauerschmidt2019}
\begin{equation}
g_n(z)=g_0(z)[\mu(z)]^n,
\label{eq:Bethe_resolvent_ansatz}
\end{equation}
where $\mu(z)$ obeys
\begin{equation}
z=\frac{1}{\mu(z)}+(q-1)\mu(z).
\end{equation}
Equivalently,
\begin{equation}
\mu(z)
=
\frac{z-\sqrt{z^2-4(q-1)}}{2(q-1)}.
\label{eq:Bethe_alpha}
\end{equation}
The branch of the square root is chosen so that
$\mu(z)\sim 1/z$ and $g_n(z)\to 0$ as $|z|\to\infty$.
Substituting Eq.~\eqref{eq:Bethe_resolvent_ansatz} into
Eq.~\eqref{eq:Bethe_resolvent_recursion_0} gives
\begin{equation}
g_0(z)=\frac{1}{z-q\mu(z)}.
\end{equation}
Thus the off-diagonal resolvent is
\begin{equation}
g_n(z)
=
\frac{\mu(z)^n}{z-q\mu(z)}.
\label{eq:Bethe_offdiagonal_resolvent}
\end{equation}
This expression is the off-diagonal analogue of the continued-fraction Green's function discussed in App.~\ref{appendix:LDOS}. 
The corresponding diagonal spectral measure is
\begin{equation}
\nu_i(\lambda)
=
-\frac{1}{\pi}
\Im g_0(\lambda+i0^+),
\label{eq:Bethe_offdiagonal_spectral_measure}
\end{equation}
which corresponds to the Kesten--McKay measure~\cite{kesten1959,mckay1981},
\begin{equation}
\nu_i(\lambda)
=
\frac{q}{2\pi}
\frac{\sqrt{4(q-1)-\lambda^2}}{q^2-\lambda^2},
\qquad
|\lambda|\leq 2\sqrt{q-1}.
\label{eq:Bethe_Kesten_McKay}
\end{equation}
Equation~\eqref{eq:Bethe_Kesten_McKay} is the analytic Bethe lattice local density of states used for the remainder of this appendix.

\subsection{Ferromagnetic solution}
\label{app:Bethe_FM}

For the Heisenberg ferromagnet, the spin-wave spectrum is obtained from Eq.~\eqref{eq:FM_solution}.
Substituting the Kesten--McKay measure in Eq.~\eqref{eq:Bethe_Kesten_McKay} into Eq.~\eqref{eq:FM_dos_from_adjacency} gives
\begin{equation}
\rho^{\rm FM}(\omega)
=
\frac{q}{2\pi |J|S}
\frac{
\sqrt{
4(q-1)-
\left(
q-\frac{\omega}{|J|S}
\right)^2
}
}{
q^2-
\left(
q-\frac{\omega}{|J|S}
\right)^2
},
\label{eq:Bethe_FM_DOS}
\end{equation}
with support
\begin{equation}
q-2\sqrt{q-1}
\leq
\frac{\omega}{|J|S}
\leq
q+2\sqrt{q-1}.
\label{eq:Bethe_FM_band_edges}
\end{equation}
The isolated uniform mode at $\lambda=q$ lies outside the thermodynamic bulk spectral measure in Eq.~\eqref{eq:Bethe_Kesten_McKay}.
It is the Goldstone mode associated with the global $\mathrm{SO}(3)$ rotation of the ferromagnetic order parameter, but it carries zero local spectral weight in the infinite Bethe lattice.
The bulk ferromagnetic gap is therefore
\begin{equation}
\Delta_{\rm FM}
=
|J|S\left(q-2\sqrt{q-1}\right).
\label{eq:Bethe_FM_gap}
\end{equation}
This is the $p=\infty$ value shown by the dashed horizontal lines in the ferromagnetic panel of Fig.~\ref{fig:gap}, while Eq.~\eqref{eq:Bethe_FM_DOS} corresponds to the $p=\infty$ row of Fig.~\ref{fig:FM_DOS}.

Since the fully polarized ferromagnetic state is an exact eigenstate, there is no zero-point correction to the ground state energy or the ordered moment.
The finite-temperature spin reduction is
\begin{equation}
\delta m(T)
=
\int_{-2\sqrt{q-1}}^{2\sqrt{q-1}}
\dd\lambda \,
\nu_i(\lambda)\,
n_{\rm B}
\left(
\frac{|J|S(q-\lambda)}{T}
\right).
\label{eq:Bethe_FM_thermal_depletion}
\end{equation}
The spin-wave estimate of the transition temperature is then obtained from
\begin{equation}
S
=
\delta m(T_{\rm c}^*).
\label{eq:Bethe_FM_Tc_equation}
\end{equation}
Solving Eq.~\eqref{eq:Bethe_FM_Tc_equation} gives the $p=\infty$ reference values in the ferromagnetic panel of Fig.~\ref{fig:Tc}.
In the large-$q$ limit, the Bethe lattice adjacency density of states becomes sharply concentrated on the scale of $q$, and the order parameter reduction becomes a single Bose factor,
\begin{equation}
\delta m(T)
\to
n_{\rm B}\left(\frac{|J|Sq}{T}\right),
\qquad
q\to\infty.
\label{eq:Bethe_FM_large_q_depletion}
\end{equation}
Thus
\begin{equation}
\frac{T_{\rm c}^*}{|J|Sq}
\to
\frac{1}{\ln\left(1+\frac{1}{S}\right)},
\qquad
q\to\infty.
\label{eq:Bethe_FM_large_q_Tc}
\end{equation}
This large-$q$ scale is the horizontal reference used in Fig.~\ref{fig:Tc}.

The static local transverse susceptibility follows from Eq.~\eqref{eq:real_chi_from_dos}.
For the ferromagnet, $u=1$ and $v=0$, so
\begin{align}
\Re\chi_{ii}^{+-}(0)
&=
-2
\int_0^\infty
\dd\omega\,
\frac{\rho^{\rm FM}(\omega)}{\omega}
\nonumber \\
&=
-\frac{2}{|J|S}
\int_{-2\sqrt{q-1}}^{2\sqrt{q-1}}
\dd\lambda\,
\frac{\nu_i(\lambda)}{q-\lambda}.
\label{eq:Bethe_FM_static_chi_integral}
\end{align}
Using Eq.~\eqref{eq:Bethe_resolvent_definition} at $z=q$ gives
\begin{equation}
\int
\dd\lambda\,
\frac{\nu_i(\lambda)}{q-\lambda}
=
g_0(q)
=
\frac{q-1}{q(q-2)}.
\end{equation}
Therefore
\begin{equation}
\Re\chi_{ii}^{+-}(0)
=
-\frac{2(q-1)}{|J|S\,q(q-2)}.
\label{eq:Bethe_FM_static_chi}
\end{equation}

\subsection{Antiferromagnetic solution}
\label{app:Bethe_AFM}

For the antiferromagnet $J>0$, the spin-wave spectrum is obtained from Eq.~\eqref{eq:AFM_solution}.
Using Eq.~\eqref{eq:AFM_dos_from_adjacency} with the Kesten--McKay measure in Eq.~\eqref{eq:Bethe_Kesten_McKay} gives
\begin{equation}
\rho^{\rm AFM}(\omega)
=
\frac{q}{\pi\omega}
\sqrt{
\frac{
\omega^2-(JS)^2(q-2)^2
}{
(JS)^2 q^2-\omega^2
}
},
\label{eq:Bethe_AFM_DOS}
\end{equation}
with support
\begin{equation}
q-2
\leq
\frac{\omega}{JS}
\leq
q .
\label{eq:Bethe_AFM_band_edges}
\end{equation}
The corresponding bulk antiferromagnetic gap is
\begin{equation}
\Delta_{\rm AFM}
=
JS(q-2).
\label{eq:Bethe_AFM_gap}
\end{equation}
Equation~\eqref{eq:Bethe_AFM_gap} gives the $p=\infty$ dashed horizontal lines in the antiferromagnetic panel of Fig.~\ref{fig:gap}, while Eq.~\eqref{eq:Bethe_AFM_DOS} gives the dashed curves in Fig.~\ref{fig:AFM_LDOS}.

The zero-temperature correction to the staggered magnetization is obtained from Eq.~\eqref{eq:AFM_zero_point_magnetization}.
Using Eq.~\eqref{eq:AFM_solution} and the bipartite symmetry of $\nu_i(\lambda)$, we find
\begin{align}
\delta m_s(0)
&=
-\frac{1}{2}
+
\frac{q^2}{2\pi}
\int_0^{2\sqrt{q-1}}
\dd\lambda\,
\frac{
\sqrt{4(q-1)-\lambda^2}
}{
\left(q^2-\lambda^2\right)^{3/2}
} .
\label{eq:Bethe_AFM_delta_m_integral}
\end{align}
Writing Eq.~\eqref{eq:Bethe_AFM_delta_m_integral} in terms of the complete elliptic integrals of the first and second kind, $K(m)$ and $E(m)$, leads to the closed form
\begin{equation}
\delta m_s(0)
=
-\frac{1}{2}
+
\frac{q}{2\pi}
\left[
K\!\left(\frac{4(q-1)}{q^2}\right)-E\!\left(\frac{4(q-1)}{q^2}\right)
\right].
\label{eq:Bethe_AFM_delta_m_elliptic}
\end{equation}
This expression gives the $p=\infty$ Bethe lattice reference values for the zero-temperature magnetization correction shown in Fig.~\ref{fig:AFM_magnetization_T0}.

The $O(S)$ zero-point contribution to the ground-state energy per spin, defined in Eq.~\eqref{eq:AFM_zero_point_energy}, is
\begin{align}
\varepsilon_{\rm qu}(q)
=
-\frac{JSq}{2}
+
\frac{JS}{2\pi}
\Biggl[
q^2 &E\!\left(\frac{4(q-1)}{q^2}\right) \nonumber \\
&-
(q-2)^2 K\!\left(\frac{4(q-1)}{q^2}\right)
\Biggr].
\label{eq:Bethe_AFM_zero_point_energy}
\end{align}
This is the $p=\infty$ reference value shown by the dashed lines in Fig.~\ref{fig:AFM_ZeroPoint}.
The first term in Eq.~\eqref{eq:Bethe_AFM_zero_point_energy} is the $O(S)$ part of the classical contribution in the convention of Eq.~\eqref{eq:AFM_zero_point_energy}, while the second term is the zero-point magnon contribution.

At finite temperature, the thermal correction to the ordered moment can be written as
\begin{align}
\delta &m_s(T)-\delta m_s(0) \nonumber \\
&=
\frac{4q^2(q-1)}{\pi}
\int_0^{\pi/2}
\dd\theta\,
\frac{
\cos^2\theta
}{
\left[
q^2-4(q-1)\sin^2\theta
\right]^{3/2}
} \nonumber \\
&\qquad\qquad\qquad\quad\times
n_{\rm B}
\left(
\frac{JS}{T}
\sqrt{q^2-4(q-1)\sin^2\theta}
\right).
\label{eq:Bethe_AFM_thermal_theta}
\end{align}
Equivalently, defining $w_0 \equiv (q-2)/q$ and changing variables to
\begin{equation}
w
=
\frac{1}{q}
\sqrt{
q^2-4(q-1)\sin^2\theta
},
\end{equation}
one obtains
\begin{equation}
\delta m_s(T)-\delta m_s(0)
=
\frac{q}{\pi}
\int_{w_0}^{1}
\dd w\,
\frac{
\sqrt{w^2-w_0^2}
}{
w^2\sqrt{1-w^2}
}\;
n_{\rm B}
\left(
\frac{JSq\,w}{T}
\right).
\label{eq:Bethe_AFM_thermal_u}
\end{equation}
The spin-wave estimate of the transition temperature is obtained by solving
\begin{equation}
S
=
\delta m_s(T_{\rm c}^*).
\label{eq:Bethe_AFM_Tc_equation}
\end{equation}
Equation~\eqref{eq:Bethe_AFM_Tc_equation} gives the $p=\infty$ reference values in the antiferromagnetic panel of Fig.~\ref{fig:Tc}.

The large-$q$ limit of Eq.~\eqref{eq:Bethe_AFM_thermal_u} is particularly simple.
Set $w=1-x/q$.
To leading order in $1/q$,
\begin{equation}
\frac{
\sqrt{w^2-w_0^2}
}{
w^2\sqrt{1-w^2}
}
\to
\sqrt{\frac{2-x}{x}},
\qquad
0\leq x\leq 2.
\end{equation}
The Bose factor becomes independent of $x$ at leading order,
\begin{equation}
n_{\rm B}
\left(
\frac{JSq\,w}{T}
\right)
=
n_{\rm B}
\left(
\frac{JSq}{T}
\right)
+
O \left(q^{-1} \right).
\end{equation}
Since
\begin{equation}
\frac{1}{\pi}
\int_0^2
\dd x\,
\sqrt{\frac{2-x}{x}}
=
1,
\end{equation}
the thermal correction reduces to
\begin{equation}
\delta m_s(T)-\delta m_s(0)
\to
n_{\rm B}
\left(
\frac{JSq}{T}
\right),
\qquad
q\to\infty.
\label{eq:Bethe_AFM_large_q_depletion}
\end{equation}
Since $\delta m_s(0)=O(q^{-1})$, the large-$q$ transition estimate satisfies
\begin{equation}
\frac{T_{\rm c}^*}{JSq}
\to
\frac{1}{\ln\left(1+\frac{1}{S}\right)},
\qquad
q\to\infty.
\label{eq:Bethe_AFM_large_q_Tc}
\end{equation}
This is the same large-$q$ scale obtained for the Heisenberg ferromagnet [see Eq.~(\ref{eq:Bethe_FM_large_q_Tc})].

Finally, the static local transverse susceptibility follows from Eq.~\eqref{eq:real_chi_from_dos}.
For the bipartite antiferromagnet,
\begin{equation}
u_\lambda^2+v_\lambda^2
=
\frac{q}{\sqrt{q^2-\lambda^2}},
\end{equation}
thus
\begin{align}
\Re\chi_{ii}^{+-}(0)
&=
-2
\int_0^\infty
\dd\omega\,
\rho^{\rm AFM}(\omega)
\frac{
u^2(\omega)+v^2(\omega)
}{\omega}
\nonumber \\
&=
-\frac{2q}{JS}
\int_{-2\sqrt{q-1}}^{2\sqrt{q-1}}
\dd\lambda \,
\frac{
\nu_i(\lambda)
}{
q^2-\lambda^2
} .
\label{eq:Bethe_AFM_static_chi_integral}
\end{align}
Using the bipartite symmetry of $\nu_i(\lambda)$,
\begin{equation}
\int
\dd\lambda \,
\frac{\nu_i(\lambda)}{q^2-\lambda^2}
=
\frac{g_0(q)}{q}
=
\frac{q-1}{q^2(q-2)}.
\end{equation}
Therefore
\begin{equation}
\Re\chi_{ii}^{+-}(0)
=
-\frac{2(q-1)}{JS\,q(q-2)}.
\label{eq:Bethe_AFM_static_chi}
\end{equation}
When measuring energies in the corresponding exchange scale, Eqs.~\eqref{eq:Bethe_FM_static_chi} and~\eqref{eq:Bethe_AFM_static_chi} are identical.
This is the $p=\infty$ limit of the equality between ferromagnetic and antiferromagnetic static local susceptibilities shown in Fig.~\ref{fig:ReChi0}.

\subsection{Transverse correlations of the bipartite antiferromagnet}
\label{app:Bethe_correlation}

We now derive the long-distance form of the equal-time transverse correlation function for the bipartite Heisenberg antiferromagnet on the $q$-regular Bethe lattice.
We denote the graph distance between two sites $i$ and $j$ by the integer
\begin{equation}
n \equiv \frac{r_{ij}}{a}.
\end{equation}
Because the Bethe lattice is bipartite, the parity of $n$ determines whether $i$ and $j$ lie on the same or opposite sublattices.

We express the equal-time transverse correlation function in terms of the off-diagonal spectral measure derived in App.~\ref{app:Bethe_green_function}.
The antiferromagnetic spin-wave Hamiltonian is diagonalized in Sec.~\ref{subsec:HeisenbergModel}, with
\begin{equation}
\omega_\lambda
=
JS\sqrt{q^2-\lambda^2},
\end{equation}
and Bogoliubov coefficients
\begin{equation}
u_\lambda^2+v_\lambda^2
=
\frac{JSq}{\omega_\lambda}
=
\frac{q}{\sqrt{q^2-\lambda^2}},
\qquad
2u_\lambda v_\lambda
=
\frac{\lambda}{\sqrt{q^2-\lambda^2}}.
\label{eq:Bethe_coherence_identities}
\end{equation}
The transverse correlator defined in Eq.~\eqref{eq:AFM_transverse_correlation} alternates in sign with sublattice parity.
In the locally rotated spin frame used in Sec.~\ref{subsec:HeisenbergModel}, same-sublattice correlations ($n$ even) involve the normal bosonic contraction, while opposite-sublattice correlations ($n$ odd) involve the anomalous contraction.
Without loss of generality, we assume that $n$ is even, as one can easily modify the following calculation to show that $|C_\perp(na)|$ has the same asymptotic form for odd $n$.

Using Eq.~\eqref{eq:Bethe_coherence_identities}, 
\begin{equation}
C_\perp(na)
=
\left[
\frac{q}{\sqrt{q^2-\vb*{A}^2}}
\right]_{ij},
\qquad n\ \mathrm{even}.
\label{eq:Bethe_correlation_operator_form}
\end{equation}
Equivalently, in spectral form,
\begin{equation}
C_\perp(na)
=
-\frac{q}{\pi}\int_{-2\sqrt{q-1}}^{2\sqrt{q-1}}
\dd\lambda\,
\frac{\Im g_n(\lambda+i0^+)}{\sqrt{q^2-\lambda^2}}.
\label{eq:Bethe_correlation_spectral_form}
\end{equation}
The isolated antiferromagnetic zero modes at $\lambda=\pm q$ are not part of the normalizable bulk spectrum of the infinite Bethe lattice.
Equivalently, on finite regular bipartite graphs their local weight is $O(1/N)$, so they are excluded from the thermodynamic bulk correlator, as in Sec.~\ref{subsec:transverse_correlations}.

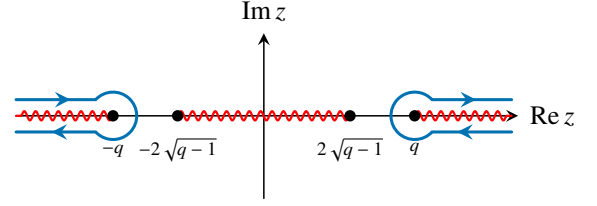
\begin{figure}[t]
\centering
\definecolor{contourcol}{RGB}{0,114,178}
\resizebox{0.9\columnwidth}{!}{
\begin{tikzpicture}[line cap=round,>=stealth,contour/.style={contourcol,thick,postaction={decorate,decoration={markings,mark=at position 0.20 with {\arrow{>}},mark=at position 0.87 with {\arrow{>}}}}}]
\def\qpt{1.4}
\def\qsq{0.8}
\def\axlen{2.3}
\def\eps{0.15}
\def\rcap{0.22}
\pgfmathsetmacro{\alphacap}{asin(\eps/\rcap)}
\pgfmathsetmacro{\xcapL}{-\qpt - sqrt(\rcap*\rcap - \eps*\eps)}
\pgfmathsetmacro{\xcapR}{\qpt + sqrt(\rcap*\rcap - \eps*\eps)}

\draw[->] (-\axlen,0) -- (\axlen+0.06,0)
    node[right] {\scriptsize$\Re z$};

\draw[->] (0,-\axlen/3) -- (0,\axlen/3)
    node[above] {\scriptsize$\Im z$};

\draw[contour]
    (-\axlen,\eps)
    -- (\xcapL,\eps)
    arc[
        start angle={180-\alphacap},
        end angle={-180+\alphacap},
        radius=\rcap
    ]
    -- (-\axlen,-\eps);

\draw[contour]
    (\axlen,-\eps)
    -- (\xcapR,-\eps)
    arc[
        start angle={360-\alphacap},
        end angle={\alphacap},
        radius=\rcap
    ]
    -- (\axlen,\eps);

\draw[red,semithick,decorate,
    decoration={snake, amplitude=1.2pt, segment length=3pt}
] (-\qpt,0) -- (-\axlen,0);

\draw[red,semithick,decorate,
    decoration={snake, amplitude=1.2pt, segment length=3pt}
] (\qpt,0) -- (\axlen,0);

\draw[red,semithick,decorate,
    decoration={snake, amplitude=1.2pt, segment length=3pt}
] (-\qsq,0) -- (\qsq,0);

\fill (-\qpt,0) circle (1.5pt);
\fill (\qpt,0) circle (1.5pt);
\fill (-\qsq,0) circle (1.5pt);
\fill (\qsq,0) circle (1.5pt);

\node[below=4pt] at (-\qpt,0) {\tiny$-q$};
\node[below=4pt] at (\qpt,0) {\tiny$q$};
\node[below=2.5pt] at (\qsq,0) {\tiny$2\sqrt{q-1}$};
\node[below=2.5pt] at (-\qsq,0) {\tiny$-2\sqrt{q-1}$};
\end{tikzpicture}
}
\caption{
Hankel contours used to extract the branch-point contribution to Eq.~\eqref{eq:Bethe_correlation_contour}.
The central branch cut is the Bethe lattice adjacency spectrum, while the outer cuts begin at $z=\pm q$ and arise from the factor $\sqrt{q^2-z^2}$ in the antiferromagnetic correlation function.
}
\label{fig:Bethe_Hankel_contours}
\end{figure}

We now extract the large-$n$ asymptotics of Eq.~\eqref{eq:Bethe_correlation_spectral_form} by writing it as a contour integral,
\begin{equation}
C_\perp(na)
=
\frac{q}{2\pi i}
\oint_{\Gamma}
\dd z\,
\frac{g_n(z)}{\sqrt{q^2-z^2}},
\label{eq:Bethe_correlation_contour}
\end{equation}
where the initial contour $\Gamma$ encloses the Bethe lattice adjacency spectrum
$\left[ -2\sqrt{q-1},2\sqrt{q-1} \right]$.
The square root is chosen to be positive for $z\in(-q,q)$, with branch cuts extending from $z=q$ to $+\infty$ and from $z=-q$ to $-\infty$.
The integrand of Eq.~\eqref{eq:Bethe_correlation_contour} is analytic in the region between the central Bethe lattice continuum and these outer branch cuts.
Therefore, the contour $\Gamma$ can be deformed away from the adjacency spectrum into two Hankel contours encircling the square root branch cuts that emanate from $z=\pm q$, as depicted in Fig.~\ref{fig:Bethe_Hankel_contours}.
The leading large-distance behavior is then controlled by the branch points at $z=\pm q$.

Consider first the branch cut beginning at $z=q$.
Write $z=q+s$, with $s>0$.
Near this branch point,
\begin{align}
\mu(q+s)
&=
\frac{1}{q-1}
\left[
1-\frac{s}{q-2}+O(s^2)
\right],
\label{eq:Bethe_alpha_near_q} \\
\frac{1}{z-q\mu(z)}
&=
\frac{q-1}{q(q-2)}
+O(s),
\label{eq:Bethe_G_near_q}
\end{align}
and
\begin{equation} \begin{split} &\frac{q}{2\pi i} \left[ \frac{1} {\sqrt{q^2-(q+s+i0^+)^2}} - \frac{1} {\sqrt{q^2-(q+s-i0^+)^2}} \right] \\ &\qquad = \frac{q} {\pi\sqrt{(q+s)^2-q^2}} = \frac{1}{\pi}\sqrt{\frac{q}{2s}} \left[1+O(s)\right], \qquad s>0 . \end{split} 
\label{eq:Bethe_branch_discontinuity_q} 
\end{equation}
Combining Eqs.~\eqref{eq:Bethe_alpha_near_q}--\eqref{eq:Bethe_branch_discontinuity_q}, the contribution from the $z=q$ branch point is
\begin{align}
I_q(n)
&=
\frac{q-1}{q(q-2)}
\frac{1}{(q-1)^n}
\frac{1}{\pi}\sqrt{\frac{q}{2}}
\nonumber \\
&\qquad\qquad\times\int_0^\infty
\dd s\,
s^{-1/2}
e^{-ns/(q-2)}
\left[1+O(s)\right] \nonumber \\
&=
\frac{q-1}{\sqrt{2\pi q(q-2)}}
\frac{(q-1)^{-n}}{\sqrt{n}}
+
O\!\left(
\frac{(q-1)^{-n}}{n^{3/2}}
\right).
\label{eq:Bethe_q_branch_contribution}
\end{align}
The branch point at $z=-q$ gives an identical contribution to the staggered correlator.
Adding the two branch-point contributions gives the leading long-distance correlator
\begin{equation}
C_\perp(na)
=
(q-1)
\sqrt{
\frac{2}{\pi q(q-2)}
}
\frac{(q-1)^{-n}}{\sqrt{n}}
+
O\!\left(
\frac{(q-1)^{-n}}{n^{3/2}}
\right)
.
\label{eq:Bethe_transverse_correlation_asymptotic_n}
\end{equation}
Restoring the graph distance $r=na$, this becomes
\begin{equation}
|C_\perp(r)|
=
\sqrt{
\frac{2(q-1)^2}{\pi q(q-2)}
}
\sqrt{\frac{a}{r}}
\exp\!\left(-\frac{r}{\xi_\perp}\right)
+
O\!\left[
\left(\frac{a}{r}\right)^{3/2}
\exp\!\left(-\frac{r}{\xi_\perp}\right)
\right],
\label{eq:Bethe_transverse_correlation_asymptotic_r}
\end{equation}
with transverse correlation length
\begin{equation}
\xi_\perp
=
\frac{a}{\ln(q-1)} .
\label{eq:Bethe_transverse_correlation_length}
\end{equation}
Equation~\eqref{eq:Bethe_transverse_correlation_asymptotic_r} is the exact leading long-distance form of the equal-time transverse correlation function on the $q$-regular Bethe lattice within linear spin-wave theory.
It has the same functional form used in Eq.~\eqref{eq:AFM_hyperbolic_transverse_fit}, but with a correlation length fixed exactly by the Bethe lattice shell-growth rate,
\begin{equation}
\lambda_{\infty,q}=q-1.
\end{equation}
Thus the Bethe lattice result saturates the inverse-volume-entropy relation
\begin{equation}
\xi_\perp=\frac{a}{\ln\lambda_{\infty,q}},
\end{equation}
which motivates our conjecture in Eq.~\eqref{eq:correlation_length_entropy_conjecture} for general bipartite hyperbolic $\{p,q\}$ lattices.

\section{Continuum limit}\label{sec:continuum}
\subsection{Effective field theory of the hyperbolic antiferromagnet}
\label{app:AFM_continuum}

Here, we derive the effective Goldstone action in the continuum limit for an antiferromagnetic Heisenberg model in $d$-dimensional hyperbolic space.
We use Poincar\'e ball coordinates $\vb*{x}\in \mathbb{R}^d$ with $|\vb*{x}|<1$, equipped with the metric
\begin{equation}
    \dd s^2 = \frac{4}{|\kappa|}\frac{|\dd \vb*{x}|^2}{(1-|\vb*{x}|^2)^2}.
\end{equation}
Consider the partition function of the Heisenberg model in the basis of spin-coherent states,
\begin{equation}
    Z = \int \mathcal{D}[\vb*{\Omega}(\tau)]
    \;\delta \left(|\vb*{\Omega}(\tau)|-1 \right)
    \; e^{-S[\vb*{\Omega}(\tau)]},
\end{equation}
with imaginary-time action (and the imaginary time $\tau$ made implicit),
\begin{equation}
\begin{split}
    S[\vb*{\Omega}] 
    =
    \int \dd{\tau}
    \left[
    -iS \sum_i \vb*{A}(\vb*{\Omega}_i) \cdot \partial_\tau \vb*{\Omega}_i 
    + 
    JS^2 \sum_{\expval{i,j}} \vb*{\Omega}_i \cdot \vb*{\Omega}_j
    \right],
\end{split}
\end{equation}
and $\vb*{A}(\vb*{\Omega}_i(\tau)) \equiv \tan\left(\frac{\theta_i(\tau)}{2}\right) \vu*{\phi}_i(\tau)$.
For the antiferromagnet, expand $\vb*{\Omega}_i$ in terms of longitudinal and transverse fields~\cite{auerbach1994},
\begin{equation}
    \vb*{\Omega}_i(\tau) = \eta_i \vb*{n}(\vb*{x}_i)
    \sqrt{1- \frac{|\vb*{L}(\vb*{x}_i)|^2}{S^2}} + 
    \frac{\vb*{L}(\vb*{x}_i)}{S},
\end{equation}
where $\eta_i = \pm 1$ has opposite signs on the two sublattices.
The constraints in this case are
\begin{equation}
    |\vb*{n}(\vb*{x}_i)| = 1, \qquad \vb*{n}(\vb*{x}_i) \cdot \vb*{L}(\vb*{x}_i) = 0.
\end{equation}
We may expand the exchange terms to quadratic order in $|\vb*{L}|$ as 
\begin{equation}
\begin{split}
    \vb*{\Omega}&(\vb*{x}_i) \cdot \vb*{\Omega}(\vb*{x}_j)
    \approx
    \eta_i \eta_j
    - \frac{1}{2}\eta_i\eta_j
    \left|
    \vb*{n}(\vb*{x}_i)-\vb*{n}(\vb*{x}_j)
    \right|^2 \\
    &-\frac{\eta_i\eta_j}{2S^2}
    \left[
    \left|\vb*{L}(\vb*{x}_i)\right|^2
    +
    \left|\vb*{L}(\vb*{x}_j)\right|^2
    \right]
    +\frac{1}{S^2}
    \vb*{L}(\vb*{x}_i)\cdot \vb*{L}(\vb*{x}_j) \\
    &+\frac{\eta_i}{S}
    \vb*{n}(\vb*{x}_i)\cdot\vb*{L}(\vb*{x}_j)
    +\frac{\eta_j}{S}
    \vb*{L}(\vb*{x}_i)\cdot\vb*{n}(\vb*{x}_j)
    +\cdots .
\end{split}
\end{equation}
Defining the displacement vector $\vb*{\delta}_{ij} = \vb*{x}_j-\vb*{x}_i$, we have
\begin{align}
    \vb*{n}(\vb*{x}_i) 
    &\approx 
    \vb*{n}(\vb*{x}_j) 
    -\delta_{ij}^a \nabla_a \vb*{n}(\vb*{x}_j)
    +\frac{1}{2} \delta_{ij}^a \delta_{ij}^b \nabla_a \nabla_b \vb*{n}(\vb*{x}_j), \\
    \vb*{L}(\vb*{x}_i) 
    &\approx 
    \vb*{L}(\vb*{x}_j) 
    -\delta_{ij}^a \nabla_a \vb*{L}(\vb*{x}_j)
    +\frac{1}{2} \delta_{ij}^a \delta_{ij}^b \nabla_a \nabla_b \vb*{L}(\vb*{x}_j).
\end{align}
Keeping terms to quadratic order in the canting field $\vb*{L}$ and to leading nontrivial order in the gradient expansion, $O(\nabla^2)$, we obtain
\begin{subequations}
\begin{align}
&\sum_{\expval{i,j}}
\eta_i \eta_j
\left|
\vb*{n}(\vb*{x}_i)-\vb*{n}(\vb*{x}_j)
\right|^2
\notag \\
&\quad \approx
\frac{1}{2}\sum_j \eta_j \sum_{i\in \mathcal{N}(j)} \eta_i
\delta_{ij}^a \delta_{ij}^b
\nabla_a \vb*{n}(\vb*{x}_j)
\cdot
\nabla_b \vb*{n}(\vb*{x}_j)
\notag \\
&\quad =
-
\frac{qa^2}{2d} \sum_j
g^{ab}(\vb*{x}_j)
\nabla_a \vb*{n}(\vb*{x}_j)
\cdot
\nabla_b \vb*{n}(\vb*{x}_j),
\end{align}

\begin{align}
\sum_{\expval{i,j}} \eta_i \eta_j
\left[
\left|\vb*{L}(\vb*{x}_i)\right|^2
+
\left|\vb*{L}(\vb*{x}_j)\right|^2
\right]
=
-q
\sum_j
\left|\vb*{L}(\vb*{x}_j)\right|^2,
\end{align}

\begin{align}
&\sum_{\expval{i,j}}
\vb*{L}(\vb*{x}_i)
\cdot
\vb*{L}(\vb*{x}_j)
\notag \\
&\quad \approx
\frac{1}{2}\sum_j
\vb*{L}(\vb*{x}_j)
\cdot
\biggl[
q \vb*{L}(\vb*{x}_j)
+
\frac{1}{2} \sum_{i\in \mathcal{N}(j)}
\delta_{ij}^a \delta_{ij}^b
\nabla_a \nabla_b
\vb*{L}(\vb*{x}_j)
\biggr]
\notag \\
&\quad =
\frac{q}{2}\sum_j
|\vb*{L}(\vb*{x}_j)|^2
+
\frac{qa^2}{4d} \sum_j
\vb*{L}(\vb*{x}_j)
\cdot
\Delta_g \vb*{L}(\vb*{x}_j),
\end{align}

\begin{align}
&\sum_{\expval{i,j}}
\left[
\eta_i
\vb*{n}(\vb*{x}_i)
\cdot
\vb*{L}(\vb*{x}_j)
+
\eta_j
\vb*{L}(\vb*{x}_i)
\cdot
\vb*{n}(\vb*{x}_j)
\right]
\notag \\
&\quad \approx
-\frac{1}{2} \sum_j
\eta_j
\vb*{L}(\vb*{x}_j)
\cdot
\left[
\sum_{i\in \mathcal{N}(j)}
\delta_{ij}^a \delta_{ij}^b
\nabla_a \nabla_b
\vb*{n}(\vb*{x}_j)
\right]
\notag \\
&\quad =
-
\frac{qa^2}{2d} \sum_j
\eta_j
\vb*{L}(\vb*{x}_j)
\cdot
\Delta_g \vb*{n}(\vb*{x}_j),
\end{align}
\end{subequations}
where we have used the long-wavelength identity
\begin{equation}
    \sum_{i\in \mathcal{N}(j)} 
    \delta_{ij}^a \delta_{ij}^b \simeq \frac{qa^2}{d}g^{ab}(\vb*{x}_j),
\end{equation}
and $\mathcal{N}(j)$ denotes the set of all nearest neighbors of site $j$.
Here, we have retained only the leading local part of the massive canting field $\vb*{L}(\vb*{x},\tau)$ and neglected its derivative-dependent corrections.
These effects are subleading in the low-energy, long-wavelength, large-$S$ expansion and can be absorbed into renormalized parameters of the effective theory.
Similarly, the Berry phase is expanded as
\begin{align}
    \vb*{A}(\vb*{\Omega}_i(\tau)) &\cdot \partial_\tau \vb*{\Omega}_i (\tau)
    \approx 
    \eta_i \vb*{A}(\vb*{n}(\vb*{x}_i)) \cdot \partial_\tau \vb*{n}(\vb*{x}_i) \nonumber \\
    &- \frac{1}{S} \vb*{L}(\vb*{x}_i) \cdot \left(\vb*{n}(\vb*{x}_i) \times \partial_\tau \vb*{n}(\vb*{x}_i)\right)
    + O\left(\frac{|\vb*{L}|^2}{S^2}\right).
\end{align}
We next take the continuum limit $\sum_j \to V^{-1}_v\int \dd[d]{x} \sqrt{|g|}$, with $V_v$ the hyperbolic volume per lattice vertex. 
For a $\{p,q\}$ lattice in dimension $d=2$, we have $V_v = \frac{\pi}{|\kappa|p} \left[(p-2)(q-2)-4\right]$.
The effective imaginary-time action is then
\begin{align}
    S[\vb*{n},\vb*{L}] = &S_{\rm{top}}[\vb*{n}] + \int \dd{\tau} \int \dd[d]{x}\sqrt{|g|} 
    \Biggl[
    \frac{1}{2\chi_\perp V_v^2}|\vb*{L}|^2
     \nonumber \\
    &\qquad+ \frac{i}{V_v} \vb*{L} \cdot (\vb*{n} \cross \partial_\tau \vb*{n})
    + \frac{\rho_s}{2} g^{ab}\nabla_a \vb*{n} \cdot \nabla_b \vb*{n}
    \Biggl],
\end{align}
where $\rho_s \equiv JS^2qa^2/2dV_v$ and $\chi_\perp \equiv 1/2JqV_v$.
The topological term
\begin{equation}
    S_{\rm top}[\vb*{n}]
    =
    -\frac{iS}{V_v}
    \int \dd{\tau}
    \int \dd[d]{x}\sqrt{|g|}\,
    \eta(\vb*{x})\,
    \vb*{A}(\vb*{n})\cdot\partial_\tau\vb*{n}
\end{equation}
is retained formally.
For smooth low-energy fluctuations about the ordered state, the staggered Berry phases cancel between sublattices, so $S_{\rm top}$ does not contribute to the leading Gaussian Goldstone action. 
It can, however, affect nonperturbative configurations such as instanton events~\cite{murthy1990,read1990,senthil2004}.

Integrating out the transverse field $\vb*{L}(\vb*{x},\tau)$, we arrive at the $\rm{O}(3)$ nonlinear sigma model,
\begin{equation}
    S[\vb*{n}] = \frac{\rho_s}{2}
    \int \dd{\tau} \dd[d]{x}
    \sqrt{|g|}
    \left[
    \frac{1}{c^2}\left(\partial_\tau \vb*{n}\right)^2
    +
    g^{ab}\partial_a \vb*{n} \cdot \partial_b \vb*{n}
    \right],
\end{equation}
where $c=\sqrt{\rho_s/\chi_\perp}$ is the spin-wave velocity.
Expanding about an ordered state $\vb*{n} \approx (\pi_1,\pi_2,\sqrt{1-|\vb*{\pi}|^2})$, we obtain the quadratic Goldstone action
\begin{equation}
    S[\vb*{\pi}] = \frac{\rho_s}{2}\sum_{\alpha=1}^2
    \int \dd{\tau} \dd[d]{x}
    \sqrt{|g|}
    \left[
    \frac{1}{c^2}\left(\partial_\tau \pi_\alpha \right)^2
    +
    g^{ij}\partial_i \pi_\alpha \partial_j \pi_\alpha
    \right].
\end{equation}
The (Matsubara) Green's function is 
\begin{align}
    \mathcal{G}_{\alpha\beta}(\vb*{x},i\omega_n) 
    &= \int \dd{\tau} \; e^{i\omega_n \tau}\;\expval{ T_{\tau}\;\pi_\alpha(\vb*{x},\tau) \; \pi_\beta(0,0)} \nonumber \\
    &=-\frac{c^2 \delta_{\alpha\beta}}{\rho_s} \bra{\vb*{x}}\left[ (i\omega_n)^2 + c^2 \Delta_g \right]^{-1}\ket{\vb*{x}=\vb*{0}},
\end{align}
whose kernel possesses normal modes $\varphi_{k,\vu*{u}}(\vb*{x})$ obeying the eigenvalue equation
\begin{equation}
    -c^2\Delta_g\varphi_{k,\vu*{u}} = \omega_k^2\varphi_{k,\vu*{u}},
\end{equation}
with $k\geq 0$. Here, $\Delta_g f \equiv |g|^{-1/2}\, \partial_a\left(\sqrt{|g|} \, g^{ab}\partial_b f\right)$ is the Laplace-Beltrami operator on the Poincar\'e ball. In terms of the Poincar\'e ball coordinates, the eigenfunctions are~\cite{cohl2012}
\begin{equation}
    \varphi_{k,\vu*{u}}(\vb*{x}) = 
    \frac{1}{\sqrt{2}}
    \left(
    \frac{\sqrt{|\kappa|}}{2\pi}
    \right)^{d/2}
    \left|
    \frac{\Gamma\left(\frac{d-1+ik}{2}\right)}
    {\Gamma\left(\frac{ik}{2}\right)}
    \right|
    \left(\frac{1-|\vb*{x}|^2}{|\vb*{x}-\vu*{u}|^2}\right)^{\frac{1}{2}(d-1+ik)},
\end{equation}
with corresponding eigenvalues $\omega_{k} = \frac{c}{2\ell}\sqrt{k^2+(d-1)^2}$.
Here, $\vu*{u}$ is a unit vector on the boundary of the Poincar\'e ball.
The spectral representation of the propagator is then
\begin{align}
    \mathcal{G}_{\alpha\beta}&(\vb*{x},i\omega_n) 
    =
    -\frac{c^2 \delta_{\alpha\beta}}{\rho_s} 
    \int_0^\infty \dd{k} 
    \int_{S^{d-1}} \dd{\Omega_{\vu*{u}}} \;
    \frac{\varphi^{}_{k,\vu*{u}}(\vb*{x})\varphi^*_{k,\vu*{u}}(\vb*{0})}
    {(i\omega_n)^2 - \omega_{k}^2} 
    \nonumber \\
    &=
    -\frac{c^2 \delta_{\alpha\beta}}{\rho_s\Gamma(d/2)} 
    \left(\frac{|\kappa|}{4\pi}\right)^{d/2}
    \int_0^\infty \dd{k} \;
    \left|\frac{\Gamma\left(\frac{d-1+ik}{2}\right)}
    {\Gamma\left(\frac{ik}{2}\right)}\right|^2
    \nonumber \\
    &\qquad\qquad\qquad\times
    \frac{{}_{2}F_{1}\left(\frac{d-1+ik}{2},\frac{d-1-ik}{2};\frac{d}{2};\frac{|\vb*{x}|^2}{|\vb*{x}|^2-1}\right)}
    {(i\omega_n)^2-\omega_k^2},
\label{eq:MatsubaraG}
\end{align}
where ${}_{2}F_{1}(a,b;c;z)$ is the hypergeometric function~\cite{cohl2012}.
Note that Eq.~\eqref{eq:MatsubaraG} is isotropic in the Poincar\'e ball coordinates.
The antiferromagnetic gap in the continuum limit is
\begin{equation}
\Delta_{\rm{AFM}} = \frac{c}{2}\sqrt{|\kappa|}(d-1).
\end{equation}
We can also compute the equal-time correlator
\begin{align}
    \mathcal{C}_{\alpha\beta}&(|\vb*{x}|) = S \expval{\pi_\alpha(\vb*{x},\tau=0)\pi_\beta(\vb*{0},\tau=0)} \nonumber \\
    &= TS \sum_{i\omega_n} \mathcal{G}_{\alpha \beta}(\vb*{x},i\omega_n) \nonumber \\
    &= \frac{S c^2 \delta_{\alpha\beta}}
    {2\rho_s\Gamma(d/2)} 
    \left(\frac{|\kappa|}{4\pi}\right)^{d/2} \nonumber \\
    &\quad\times\int_0^\infty \dd{k} \;
    \left|\frac{\Gamma\left(\frac{d-1+ik}{2}\right)}
    {\Gamma\left(\frac{ik}{2}\right)}\right|^2 
    \left(\frac{1+2n_{\rm B}(\omega_k/T)}
    {\omega_k}\right) \nonumber \\
    &\quad\times{{}_{2}F_{1}}\left(\frac{d-1+ik}{2},\frac{d-1-ik}{2};\frac{d}{2};\frac{|\vb*{x}|^2}{|\vb*{x}|^2-1}\right).
    \label{eq:eq_time_correlation_continuum}
\end{align}
To find the long-distance ($r \equiv 2|\kappa|^{-1/2}\; \mathrm{arctanh}(| \vb*{x} |)\to \infty$) behavior of Eq.~\eqref{eq:eq_time_correlation_continuum}, we rewrite the hypergeometric function as~\cite{NIST2010}
\begin{align}
    &{{}_{2}F_{1}}\left(a,b;c;-u \right) \nonumber \\
    &= 
    u^{-a}
    \frac{\Gamma(c)\Gamma(b-a)}{\Gamma(b)\Gamma(c-a)}
    {{}_{2}F_{1}}\left(a,1-c+a;1-b+a;-\frac{1}{u} \right) \nonumber \\
    &\;+
    u^{-b}
    \frac{\Gamma(c)\Gamma(a-b)}{\Gamma(a)\Gamma(c-b)}
    {{}_{2}F_{1}}\left(b,1-c+b;1-a+b;-\frac{1}{u} \right).
\end{align}
Then we have, as $u \to \infty$, ${{}_{2}F_{1}}\left(a,b;c;-\frac{1}{u}\right) = 1 + O(u^{-1})$.
It follows that the leading asymptotic behavior is 
\begin{align}
    &{{}_{2}F_{1}}\left(\frac{d-1+ik}{2},\frac{d-1-ik}{2};\frac{d}{2};-\sinh^2\left(\frac{1}{2}\sqrt{|\kappa|}r\right) \right) 
    \nonumber \\
    &\quad\approx
    2^{d-1}\Gamma\left(\frac{d}{2}\right)
    e^{-\frac{d-1}{2}\sqrt{|\kappa|}r}
    \left[
    \frac{2^{ik} \Gamma\left(-ik\right) e^{-i\frac{k}{2}\sqrt{|\kappa|}r}}
    {\Gamma\left(\frac{d-1-ik}{2}\right)\Gamma\left(\frac{1-ik}{2}\right)}
    +
    \textrm{c.c.}
    \right].
\end{align}
Inserting this into the integrand, we have
\begin{align}
    &\mathcal{C}_{\alpha\beta}\left( \tanh \left(\frac{1}{2}\sqrt{|\kappa|}r\right)\right) \nonumber \\
    &\approx 
    \frac{S c^2 \delta_{\alpha\beta}}
    {8\rho_s\sqrt{\pi}} 
    \left(\frac{|\kappa|}{\pi}\right)^{d/2}
    e^{-\frac{d-1}{2}\sqrt{|\kappa|}r} \nonumber \\
    &\qquad\times
    \int_0^\infty \dd{k} \;
    \left(
    \frac{1+2n_{\rm B}(\omega_k/T)}{\omega_k}
    \right)
    \left[
    \frac{\Gamma\left(\frac{d-1+ik}{2}\right) e^{-i\frac{k}{2}\sqrt{|\kappa|}r}}
    {\Gamma\left(\frac{ik}{2}\right)}
    +
    \rm{c.c.}
    \right]
    \nonumber \\
    &= 
    \frac{S c \delta_{\alpha\beta}}
    {4\rho_s} 
    \frac{|\kappa|^{(d-1)/2}}{\pi^{(d+1)/2}}
    e^{-\frac{d-1}{2}\sqrt{|\kappa|}r} \nonumber \\
    &\quad \times
    \int_{-\infty}^\infty \dd{k} \;
    \left(
    \frac{1+2n_{\rm B}(\omega_k/T)}{\sqrt{k^2+(d-1)^2}}
    \right)
    \frac{\Gamma\left(\frac{d-1+ik}{2}\right) e^{-i\frac{k}{2}\sqrt{|\kappa|}r}}
    {\Gamma\left(\frac{ik}{2}\right)}
    .
    \label{eq:contour_integral}
\end{align}
\begin{figure}[t!]
\centering
\definecolor{contourcol}{RGB}{0,114,178} 
\resizebox{0.9\columnwidth}{!}{
\begin{tikzpicture}[
    line cap=round,
    >=stealth,
    contour/.style={
        contourcol,
        thick,
        postaction={
            decorate,
            decoration={
                markings,
                mark=at position 0.09 with {\arrow{>}},
                mark=at position 0.22 with {\arrow{>}},
                mark=at position 0.41 with {\arrow{>}},
                mark=at position 0.58 with {\arrow{>}},
                mark=at position 0.723 with {\arrow{>}},
                mark=at position 0.89 with {\arrow{>}}
            }
        }
    }
]
    \def\kminus{-0.5}
    \def\axlen{2.3}
    \def\R{2.0}
    \def\eps{0.11}
    \def\rcap{0.20}
    \pgfmathsetmacro{\theta}{acos(\eps/\R)}
    \pgfmathsetmacro{\phir}{-\theta}
    \pgfmathsetmacro{\phil}{-180+\theta}
    \pgfmathsetmacro{\yint}{-sqrt(\R*\R-\eps*\eps)}
    \pgfmathsetmacro{\thetacap}{acos(\eps/\rcap)}
    \pgfmathsetmacro{\ycap}{\kminus - sqrt(\rcap*\rcap - \eps*\eps)}
    \draw[->] (-\axlen,0) -- (\axlen,0)
        node[right] {\scriptsize$\Re k$};

    \draw[->] (0,-\axlen) -- (0,\axlen/2)
        node[above] {\scriptsize$\Im k$};
    \draw[contour]
        (-\R,0)
        -- (\R,0)
        arc[start angle=0,end angle=\phir,radius=\R]
        -- (\eps,\ycap)
        arc[start angle=-\thetacap,end angle=180+\thetacap,radius=\rcap]
        -- (-\eps,\yint)
        arc[start angle=\phil,end angle=-180,radius=\R];
    \draw[
        red,
        semithick,
        decorate,
        decoration={snake, amplitude=1.2pt, segment length=3pt}
    ] (0,\kminus) -- (0,-\axlen);
    \fill (0,\kminus) circle (1.5pt);
    \node[left=4pt] at (0,\kminus) {\scriptsize$-i(d-1)$};
\end{tikzpicture}
}
\caption{Contour used to compute the integral in Eq.~\eqref{eq:contour_integral}.}
\label{fig:branch-cut}
\end{figure}
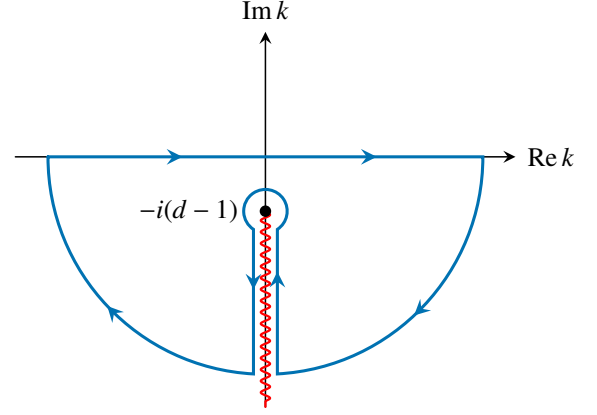
Equation~\eqref{eq:contour_integral} can be evaluated by extending the integral to a semi-circular contour enclosing the lower-half complex plane, as depicted in Fig.~\ref{fig:branch-cut}.
In this case, the only relevant singularity is the branch point, $k = -i(d-1)$.
At zero temperature, we find
\begin{align}
    C_{\perp}(r) &= \sum_{\alpha}\mathcal{C}_{\alpha\alpha}\left( \tanh \left(\frac{1}{2}\sqrt{|\kappa|}r\right)\right) \nonumber \\
    &=
    \frac{S c}
    {\rho_s} 
    \frac{|\kappa|^{(d-1)/2}}{\pi^{(d+1)/2}}
    e^{-(d-1)\sqrt{|\kappa|}r} \nonumber \\
    &\qquad\qquad\times
    \int_{0}^\infty \dd{s} \;
    \frac{\Gamma\left(d-1+\frac{s}{2}\right) e^{-\frac{s}{2}\sqrt{|\kappa|}r}}
    {\Gamma\left(\frac{d-1+s}{2}\right)\sqrt{s(s+2(d-1))}}
    \nonumber \\
    &= \frac{Sc|\kappa|^{(d-1)/2}}{\rho_s} 
    \frac{2^{d-2}\Gamma(d/2)}{\sqrt{d-1}\pi^{(d+1)/2}}
    \frac{e^{-(d-1)\sqrt{|\kappa|}r}}{\sqrt{|\kappa|^{1/2}r}} \nonumber \\
    &\qquad\qquad\qquad\qquad+
    O\left(r^{-3/2}e^{-(d-1)\sqrt{|\kappa|}r}\right).
    \label{eq:AFM_correlation_function_continuum}
\end{align}
Thus, the transverse correlations have the long-distance scaling of the form $C_{\perp}(r)\sim r^{-1/2} e^{-r/\xi_\perp}$ with transverse correlation length 
\begin{equation}
\xi_\perp = \frac{1}{(d-1)\sqrt{|\kappa|}}.
\label{eq:continuum_correlation_length}
\end{equation}
That is, the transverse correlation length is proportional to the radius of curvature $\ell = 1/\sqrt{|\kappa|}$.

\subsection{Effective field theory of the hyperbolic ferromagnet}
\label{app:FM_continuum}
The ferromagnetic continuum action follows from the same coherent-state path integral, with $\vb*{L} = \vb*{0}$ and $\eta_i=1$. 
Then the imaginary-time action is
\begin{equation}
    S[\vb*{n}] = \int \dd{\tau} \dd[d]{x}\sqrt{|g|} 
    \left[
    -i \frac{S}{V_v} \vb*{A}(\vb*{n}) \cdot \partial_\tau \vb*{n}
    +
    \frac{\rho_s}{2} g^{ab}\partial_a \vb*{n} \cdot \partial_b \vb*{n}
    \right].
\end{equation}
Expanding about the order parameter
\begin{equation}
    \vb*{n} = \left( \pi_1,\pi_2,\sqrt{1-|\vb*{\pi}|^2} \right),
\end{equation}
we obtain the quadratic action
\begin{equation}
\begin{split}
    S[\vb*{\pi}]
    &=
    \frac{S}{2V_v}
    \int \dd{\tau} \int \dd[d]{x}\sqrt{|g|} 
    \\
    &
    \times\biggl[
    -i \left( \pi_1 \partial_\tau \pi_2 - \pi_2 \partial_\tau \pi_1 \right)
    +
    \frac{ V_v \rho_s}{S} g^{ab}\partial_a \vb*{\pi} \cdot \partial_b \vb*{\pi}
    \biggr].
\end{split}
\end{equation}
We then transform $\psi \equiv (\pi_1+i\pi_2)/\sqrt{2}$ to obtain
\begin{equation}
    S[\psi] = 
    \frac{S}{2V_v}
    \int \dd{\tau} \dd[d]{x} \sqrt{|g|}\; \psi^*
    \left[
    \partial_\tau - D \Delta_g
    \right]
    \psi,
\end{equation}
where $D = |J|Sqa^2/2d$ is the ferromagnetic spin stiffness.
The gap of the Laplace-Beltrami operator is $|\kappa|(d-1)^2/4$, therefore the spin-wave gap for the ferromagnet is
\begin{equation}
    \Delta = \frac{D}{4}(d-1)^2|\kappa|.
\end{equation}

\bibliography{refs}
\end{document}